\documentclass[12pt, english,aps,pre,onecolun,superscriptaddress, floatfix, nofootinbib]{revtex4-1}

\usepackage{babel}
\usepackage[utf8]{inputenc}
\usepackage{graphicx}
\usepackage{hyperref}
\usepackage{color}
\usepackage{array}
\usepackage[countmax]{subfloat}
\usepackage{multirow}
\usepackage{epstopdf}
\usepackage{amsmath}
\usepackage{amssymb}
\usepackage{mathrsfs}
\usepackage{times}

\usepackage[format=plain,justification=centerlast,font=footnotesize,labelfont=bf, width = \textwidth]{caption}

\newcommand{\OptimizedSyn}{OS}
\newcommand {\HighSyn}{H}
\newcommand{\SmallSyn}{L}

\renewcommand{\eqref}[1]{Eq.~(\ref{#1})}

\begin{document}

\title{Scaling and optimal synergy: Two principles determining microbial growth in complex media}
\date{\today}

\author{Francesco Alessandro Massucci}
\affiliation{Departament d'Enginyeria Química, Universitat Rovira i Virgili, 43007 Tarragona, Spain}

\author{Roger Guimer\`a}
\affiliation{Instituci\'o Catalana de Recerca i Estudis Avançats (ICREA), Barcelona 08010, ES}	
\affiliation{Departament d'Enginyeria Química, Universitat Rovira i Virgili, 43007 Tarragona, Spain}

\author{Lu\'{\i}s A. Nunes Amaral}
\affiliation{Department of Chemical and Biological Engineering, Northwestern University, Evanston, IL 60208, USA}
\affiliation{Northwestern Institute on Complex Systems, (NICO) Northwestern University, Evanston, IL 60208, USA}
\affiliation{Howard Hughes Medical Institute, Northwestern University, Evanston, IL 60208, USA}

\author{Marta Sales-Pardo}
\email{marta.sales@urv.cat}
\affiliation{Departament d'Enginyeria Química, Universitat Rovira i Virgili, 43007 Tarragona, Spain}

\begin{abstract}
  High-throughput experimental techniques and bioinformatics tools make it possible to obtain reconstructions of the metabolism of microbial species. Combined with mathematical frameworks such as flux balance analysis, which assumes that nutrients are used so as to maximize growth, these reconstructions enable us to predict microbial growth.
%
  Although such predictions are generally accurate, these approaches do not give insights on how different nutrients are used to produce growth, and thus are difficult to generalize to new media or to different organisms.
  %
%
  Here, we propose a systems-level phenomenological model of metabolism inspired by the virial expansion. Our model predicts biomass production given the nutrient uptakes and a reduced set of parameters, which can be easily determined experimentally. To validate our model, we test it against in silico simulations and experimental measurements of growth, and find good agreement. From a biological point of view, our model uncovers the impact that individual nutrients and the synergistic interaction between nutrient pairs have on growth, and suggests that we can understand the growth maximization principle as the optimization of nutrient synergies.
\end{abstract}
\maketitle
\section{\label{sec:intro}Introduction}

The rapid development of high-throughput experimental techniques and bioinformatics 
tools has made it possible to obtain reliable metabolic reconstructions from genomic data in a semiautomatic fashion ~\cite{ray10, henry10,christian09,orth12}. The availability of such reconstructions makes it possible, in turn, to investigate metabolism from a systems point of view~\cite{oberhardt09}. In particular, the development of a mathematical framework to predict cellular growth based on cellular function optimization has significantly advanced our understanding of how the metabolic state of an organism will change upon modifications in the growth medium, the introduction of mutations, or the effect of stress~\cite{Varma1994, segre02,Kauffman2003,Orth2010,chang13,schuetz12,bordel13}.

Unfortunately, our ability to calculate microbial growth rates has not been paralleled by a substantial gain of insight into metabolic processes, especially for what concerns the impact of nutrients on growth. A number of mathematical models have been developed aiming at predicting microbial growth rates~\cite{bader78,egli93,kovarova-kovar98,toda03,zinn04,Boer2010}, but these models are only valid for a limited number of specific nutrients and are not easily generalizable because of the need to determine parameters empirically.

Here, we present a systems-level phenomenological model that enables us to predict growth and, at the same time, provides insights into the effective systems-level principles by which nutrients are catabolized. Our approach does {\em not} predict which nutrients will be uptaken from a given medium; rather, it predicts, {\em from} the values of the uptakes, how each nutrient will contribute to cellular growth. Despite the fact that we use flux balance analysis (FBA) to develop, justify and validate our model (and that, as we discuss later in Section~\ref{sec:scope}, FBA has well known limitations), the model is ultimately independent of FBA and of any particular metabolic reconstruction; in this sense, the model is also organism-independent.

Our approach, which is analogous to a virial expansion, reveals that cellular growth can be well-approximated by the contributions of each individual nutrient plus a synergy term that considers nutrient-pair contributions. We demonstrate that the predictions of the model are in good agreement with empirical measurements of biomass production. Moreover, our model provides novel insight into the effective contributions to growth since we can express synergy contributions as scaling functions that depend exclusively on four factors: the type of nutrients considered, the pathways that catabolize them, the ratio between their uptake fluxes, and the effective carbon content of each nutrient. Uptake fluxes are allocated among possible synergistic contributions in order to maximize synergy, thus revealing the principles of nutrient use that lead to the maximization of biomass production.

\section{\label{sec:model}Model}
Our goal is to express in closed form the steady-state growth rates
$g$ of a bacterium given the nutrient
uptakes from the external medium, without taking explicitly into account any
micro-level information about the processes occurring inside the cell.  In
\cite{Seaver2012} and \cite{beg07}, models to predict which nutrients
can produce growth and what constraints are necessary to reproduce
observed uptakes in rich media were already developed.  Here, we
consider that the real uptake fluxes of each nutrient are known and
fall within the empirical range which ensures that nutrient uptakes
can be fully catabolized~\cite{Orth2010}.

To validate our model, we use FBA predictions of biomass
production for {\it Escherichia coli} using the metabolic reconstruction
iAF1260
, which has been shown to yield a good agreement
with empirically measured growth rates \cite{Durot2009}. Note that we
focus exclusively on the use of nutrients for biomass
polymerization, discarding the role of ATP maintenance
 (see \cite{Seaver2012} and Sec.~\ref{sec:scope}). For simplicity,
we focus on nutrients that belong to one of the four main nutrient
classes: sugars, fatty acids, amino acids, and bases (see
Appendix for a complete list).

Following a virial expansion-like formulation, we hypothesize that,
given a fixed vector of nutrient uptake fluxes $\boldsymbol{\phi}$, we
can express the steady-state biomass production of an organism as
\begin{equation} \label{eq:expansion}
\begin{split}
g(\boldsymbol{\phi}) &= \sum_{i=1}^E \alpha_i (\phi_i)+ \sum_{j<k}^E \beta_{jk} \bigl(\phi_j, \phi_k\bigr) \\
&+  \sum_{i < j< k} \gamma\bigl(\phi_i, \phi_j,\phi_k \bigr)+\dots~,
\end{split}
\end{equation}
where $E$ is the number of uptakes.


A first order approximation is equivalent to considering that each single nutrient contributes independently to $g(\boldsymbol{\phi})$ as in \cite{Seaver2012}. In analogy to the ideal gas approximation, we call this model {\it idealized metabolism} (IM). 
Note that because we consider the nutrient use for stationary biomass production exclusively, 
in the presence of a single nutrient uptake ({\em i.e.} $\phi_i\neq 0$ for a single $i$ and $\phi_k\equiv0$ for $k\neq i$) the scale of our
system is precisely given by $\phi_i$. Therefore, the biomass production must be
proportional to $\phi_i$, so that $g(\phi_i)=\hat{\alpha} \phi_i$, where $ \hat{\alpha}$ is the biomass yield of nutrient $i$
~\cite{Orth2010,almaas04,Seaver2012}. For the first order terms, we thus write:
\begin{equation} \label{eq:f_1_order}
g(\boldsymbol{\phi}) =  \sum_{i=1}^E \alpha_i (\phi_i) = \sum_{i=1}^E \hat{\alpha}_i \phi_i.
\end{equation}
 We evaluate $\hat{\alpha}_i$ for each nutrient $i$ by computing the FBA biomass production $g_{\rm FBA}(\boldsymbol{\phi}^{(i)})$ allowing for a single nutrient uptake 
\begin{equation*}
\boldsymbol{\phi}^{(i)} = %
\begin{cases}
\phi_i = 1 {\rm~arb. units}, &\\
\phi_j = 0 {\rm~arb. units}\, &\forall j\neq i,
\end{cases}
\end{equation*}
where we use arbitrary units, since all fluxes are defined
 up to a multiplicative constant in the FBA problem.
Note that in \eqref{eq:f_1_order}, only purines among bases 
can be accounted for growth, 
since pyrimidines alone cannot be catabolized by {\it E. coli}
\cite{Seaver2012}. 
Previously, we found that $\hat{\alpha}_i$ is proportional to the effective number of carbons $C_i$, that is, the number of carbons that are actually catabolized \footnote{For the nutrient classes we consider, the effective carbons equal the actual carbons for all nutrients except for the bases} in each metabolite $i$ as
\begin{equation}
\label{eq:alpha_carbons}
\hat{\alpha}_i=a_c C_i,
\end{equation}
with a slope $a_c$ that is nearly insensitive to the
nutrient class $c$ (fatty acids, sugars, amino acids, 
Fig. \ref{fig:alphas}a). Here, both the vector $\boldsymbol{\hat{\alpha}}$ and
  the slopes $a_c$ are dimensionless quantities.

To assess the accuracy of the IM, we compare the predictions of the model
against FBA calculations for the growth of {\em E. coli} on random complex media with a fixed number of non-zero nutrient uptakes (Methods). Because $g$ is defined up to a multiplicative constant, the largest the total uptake, the largest the biomass production.  We thus consider complex uptake vectors normalized to 1, to mimic physiologic conditions. However, we note that we would obtain the same relative errors for a fixed number of uptakes if we considered non-normalized fluxes. 

Figure~\ref{fig:alphas}c shows that despite its simplicity, the idealized model is fairly accurate, with a relative error, $\Delta:=\frac{|g_{\rm FBA}-g(\boldsymbol{\phi})|}{g_{\rm FBA}}$, ranging from $\sim$ 0--2\% for one nutrient to 24\% for 20 uptakes. 
 Note that using \eqref{eq:alpha_carbons} to predict growth lightly overestimates single nutrient contributions to growth, as the corresponding $\Delta$ for growth on one nutrient shows. This effect however is negligible when increasing the number of uptakes above $E\geq 5$. It is also apparent that the IM systematically underestimates FBA predictions for media with $E\geq2$ nutrients, which implies that when several nutrients are present, they contribute synergistically to growth.        

\section{Results}
\subsection{\label{subsec:2nd_order}Scaling  of second order terms}

In order to capture nutrient growth synergies, we consider next the second order terms in \eqref{eq:expansion}.
Using FBA, we numerically determine $\beta_{ij}$ by setting to zero all entries of the exchange fluxes except $\phi_i$ and $\phi_j$ and computing the difference
\begin{equation}\label{eq:beta}
\begin{split}
\beta_{ij} \bigl(\phi_i, \phi_j\bigr) = g_{\rm FBA}(\boldsymbol{\phi}^{(i, j)}) - &\hat{\alpha}_i \phi_i - \hat{\alpha}_j \phi_j,\\
\end{split}
\end{equation}
where $\boldsymbol{\phi}^{(i, j)}$ is the vector $\boldsymbol{\phi}$ such that $\phi_k=0~ \forall k \neq i,j$ (Fig. \ref{3dbeta}a). 

Since there is only one output in our system (biomass), 
the scale of of $g$ is fixed by one of the uptake fluxes (for instance $\phi_j$) and the dependency on the remaining uptake fluxes can be expressed as 
dimensionless quantities, which are ratios of uptake fluxes. As a consequence, we expect $\beta$ to obey a scaling property (Fig. \ref{3dbeta}b):
\begin{equation} \label{eq:scale_beta}
\frac{1}{\phi_j}\beta_{ij}(\phi_i,\phi_j)=\beta_{ij}\left(\frac{\phi_i}{\phi_j},1\right)\equiv \beta_{ij}\left(\frac{\phi_i}{\phi_j}\right).
\end{equation}
Remarkably, we find that $\beta$ displays additional scaling properties. 
For concreteness, consider the synergy between sugars and fatty acids. We found that the $\beta$ functions for any sugar--fatty acid pair (Fig. \ref{3dbeta}c) collapse on the same curve 
when the sugar and the fatty acid uptake fluxes $\phi_i$, $\phi_j$ are rescaled with respect to the effective number of carbons $C_i$, $C_j$ of the corresponding nutrient (Fig. \ref{3dbeta}d). 
One thus has 
\begin{equation} \label{eq:betaCarbons}
\beta'_{\rm sug,f\_acid}\Bigl(\frac{\phi_i}{\phi_j}\Bigr) = \frac{1}{C_j} \beta_{ij}\left(\frac{C_i\phi_i}{C_j\phi_j}\right),
\end{equation}
so that the introduction of the rescaled $\beta'$ function allows to have a systematic description of growth only given the
 nutrient--pair classes, their carbon content and the ratio of their uptake fluxes. For each nutrient--class pair $\sigma$, $\sigma'$
  it is therefore possible to define a function $\beta'_{\sigma\sigma'}$ that displays a simple two--regime behavior  (Fig. \ref{3dbeta}d), in which one of the nutrients becomes the limiting factor in the contribution to growth.
 Considering again the case of sugars and fatty acids,
 when the ratio $C_i\phi_i/(C_j\phi_j) \to 0$ the function  $\beta'_{\rm sug,f\_acid}$ grows linearly, while when $C_i\phi_i/(C_j\phi_j) \gg 1$ it reaches a plateau. 
To capture these two regimes, we propose the generalized phenomenological model:
\begin{equation}\label{eq:tanhbeta}
\beta'_{\sigma_i \sigma_j}\Bigl(\frac{\phi_i}{\phi_j}\Bigr) = b_{\sigma_j \sigma_i}\tanh\left(\frac{b_{\sigma_i \sigma_j} C_i\phi_i}{b_{\sigma_j \sigma_i} C_j \phi_j}\right)
\end{equation}
where 
\begin{equation}\label{eq:betaPar}
\begin{split}
b_{\sigma_i\sigma_j} \equiv \lim_{\phi_i/\phi_j\to 0} \frac{\beta'(\phi_i/\phi_j)}{\phi_i/\phi_j}~,\\
b_{\sigma_j\sigma_i} \equiv \lim_{\phi_i/\phi_j\to \infty} \beta'(\phi_i/\phi_j).
\end{split}
\end{equation}
Here $\sigma_i$ and $\sigma_j$ are the classes of nutrient $i$, $j$, respectively, while $b_{\sigma_i\sigma_j}$ and $b_{\sigma_j\sigma_i}$ are dimensionless parameters, since they are defined as a flux ratio. 
These parameters can be interpreted as the limiting synergistic contribution to the biomass yield when one of the two nutrients is 
in excess of the other. In this formulation, knowing the limiting contributions is thus enough 
to compute the synergistic contribution to growth of
any sugar--fatty acid pair and  for any value of the uptake fluxes. 
For instance, the transition value $T({\rm sug},{\rm f\_acid})=b_{_{\rm f\_acid\, sug}}/b_{_{\rm sug\, f\_acid}}$ marks the relative sugar--fatty acid uptake values at which maximal synergy may be attained without waste of nutrients.


Figure \ref{fig:all_betas} shows the averaged collapsed curves for all nutrient class pairs we consider.
Our calculations indicate that \eqref{eq:tanhbeta} is a fairly good description 
for such averaged $\beta'$, although we note that for each nutrient class pair $\beta'$ has different parameters (see Table \ref{tab:summary} and Appendix  
for a summary of the averaged parameters for each one of these curves). Note that, for nutrients in the same class, it is not necessary to consider all pair permutations. One can, for instance, sort nutrients in a given class $\sigma$ by their carbon content and evaluate the parameters $b_{\sigma\sigma}$ only between pairs $i,j$ such that $C_i<C_j$. This is the approach we follow in evaluating the parameters $b_{\sigma\sigma'}$, which, as a consequence, are not symmetric when $\sigma=\sigma'$.

The phenomenological model in  \eqref{eq:tanhbeta} captures very well the behavior of $\beta'$ for 4 of the 9 cases: (fatty acid, sugar), (fatty acid, fatty acid), 
(base, sugar) and (base, base) pairs (Figs. \ref{fig:all_betas}a, d, b, and g) \footnote{Note that nutrients in the same class are ordered with their carbon content and pair permutations are not considered. Thus in $\beta'(\frac{\phi_1}{\phi_2})$, 
$\phi_1$ always corresponds to the nutrient with the smaller number of carbons. This implies that, for the $\beta'$ within the same class, the average slope and plateau values are not equal (see Table \ref{tab:summary}). We also remind that, 
in the base-base pair case, we only consider 
pairs of {\em purines} as {\it E. coli} cannot catabolize pyrimidines by themselves.}.
For the (base, fatty acid) case (Fig. \ref{fig:all_betas} e), we find that the phenomenological model in \eqref{eq:tanhbeta} does not fully capture the behavior of the averaged $\beta'$ (see Appendix). In such case we still find that $\beta'$ is roughly linear for $\phi_{\rm 1}/\phi_{\rm 2}\ll 1$ and shows a plateau 
when $\phi_{\rm 1}/\phi_{\rm 2}\gg 1 $, as predicted by \eqref{eq:tanhbeta}. However, for $C_1\phi_{\rm 1}/(C_2\phi_{\rm 2})\simeq 1$, the model overpredicts the observed synergy.
Despite this deviation, \eqref{eq:tanhbeta} is
a good trade off between model simplicity and predictive power, since 
the initial slope of $\beta'$ and the plateau value are well predicted by 
taking the average of the parameter $b$ over all nutrient pairs.


Finally, for all pairs including amino acids (Figs.~\ref{fig:all_betas} c,f, h, and i), we find that not all curves collapse into a single one. 
In particular, we see that when  $\phi_{\rm other}/\phi_{\rm a.acid}\gg 1$  ($\{{\rm other}$:
${\rm sugar}$, ${\rm f\_acid}$ ${\rm base}$, ${\rm a\_acid}\}$),  
the scaling functions reach different plateau values,
 which always lie either above
 or below a $10^{-2}$ threshold value, respectively.
 Interestingly, for interclass interactions,
 any given amino acid consistently reaches
 a plateau above or below such threshold independent
 of the other nutrient paired with it.
 We hence classify amino acids into two groups,
$\SmallSyn$~(Low synergy), $\HighSyn$~(High synergy),
 according to whether they can attain a synergy
 below or above the mentioned $10^{-2}$ threshold,
 for interclass synergies.
 For amino acid-amino acid interactions, we thus 
divide nutrients into $\HighSyn$~and $\SmallSyn$~and study
 intraclass/L-H synergies.
This allows us to find two slope and plateau values respectively,
 each related to the $\HighSyn$~or $\SmallSyn$~amino acid limiting the interaction in turn.

Using a logistic regression model, we find that the set of metabolic 
pathways in which an amino acid participates determines to which group ($\HighSyn$~or $\SmallSyn$) it belongs (see Appendix). 
By minimizing the Bayesian Information Criterion \cite{schwarz78}, we see that knowing whether the amino acid participates in the set of six pathways listed in Table \ref{tab:Aacids} is enough to correctly assign all amino acids except MD-Methionine to either group $\HighSyn$~or $\SmallSyn$.
Once the corresponding  group is known, we can use \eqref{eq:tanhbeta} to describe $\beta'$ by allowing two plateau values when the nutrient pair involves an amino acid. In this way, we can have close estimates of synergies through the function \eqref{eq:betaPar} for nutrients pairs from all classes, by only knowing their class and the pathways in which they participate.

\subsection{\label{subsec:2nd_order_synergy}Competition for synergistic potentials}

When a bacterium grows on a complex medium with $E>2$ nutrients, \eqref{eq:expansion}  yields a sum over $E(E-1)/2$ synergy contributions resulting in an overprediction of the biomass production (see Appendix). The reason for this is that resources are limited by stoichiometry, thus besides the  independent nutrient contribution to growth of each uptake $\phi_i$, resources must be distributed in some way among the $E-1$ possible synergies. Two plausible flux allocations are the following: i) an equitative distribution of all $\{\phi_i\}$ among the synergies ({\em equitative synergy model}, ES); ii) a  distribution among synergies that yields maximal synergy, which we call {\em optimal synergy model} (\OptimizedSyn). We find that while the former  underpredicts growth rates when increasing the number of uptakes, the latter yields an accurate prediction of FBA growth rates roughly independent of the number of nutrients (Fig. \ref{fig:exper_beta} and Appendix). Our results thus suggest that, phenomenologically, one can understand the growth maximization principle observed in microbes as the optimization of nutrient synergies.

The \OptimizedSyn~theory exploits the fact that synergy contributions are limited by the smallest uptake flux \eqref{eq:tanhbeta}, so that only the nutrients in excess can be used in other synergies. In order to maximize the overall synergy, we hypothesize that an optimal allocation of nutrients is adopted to produce the largest pair--synergies. We thus rank nutrient--pair synergies and add up to the total synergy each contribution. After each addition, the fluxes of the pair are rescaled such that the limiting one is not considered further, while the nutrient in excess can contribute to other synergies with the fraction of uptake not invested yet (Methods).

In a complex growth medium with $E$ non-zero nutrient uptakes, we thus express the \OptimizedSyn~growth rate as follows:
\begin{equation} \label{eq:secondorder}
\begin{split}
g(\boldsymbol{\phi}) &= \sum_{\ell=1}^E \hat{\alpha}_{\sigma_{\ell}} \phi_\ell\\
&+ \sum_{(\kappa , \jmath)=1}^{P} b_{\sigma_{\kappa} \sigma_{\jmath}} C_{\kappa} q^{r_{\kappa\jmath}}_{\kappa}\phi_{\kappa} 
\tanh \Biggl(\frac{b_{\sigma_{\jmath}\sigma_{\kappa} }q^{r_{\kappa\jmath}}_{\kappa}\phi_\kappa C_\kappa}{ b_{\sigma_{\kappa} \sigma_{\jmath}}q^{r_{\kappa\jmath}}_{\jmath}\phi_\jmath C_\jmath}\Biggr),
\end{split}
\end{equation}
where the second sum runs over the $P=E(E-1)/2$ ranked pairs of nutrients, $r_{\kappa\jmath}$ is the ranking of the nutrient pair synergy $(\kappa,\jmath)$, and $q^{r_{\kappa\jmath}}_{\kappa}\in [0,1]$ indicates the fraction of uptake flux $\phi_{\kappa}$ yet to be allocated to this contribution. As before, $C_\ell$ is the effective number of carbons of nutrient $\ell$ and $\sigma_\ell$ is the nutrient class to which nutrient $\ell$ belongs, and coefficients $b$ have been reported in Table \ref{tab:summary}. The yields $\hat{\alpha}_{\sigma_{\ell}}$ can either be directly evaluated for each nutrient, or computed as in equation \eqref{eq:alpha_carbons}, with parameters $a$ reported in Fig. \ref{fig:alphas} b. Note that, when available it is preferable to use the exact $\hat{\alpha}$ when dealing with less than 4 nutrients, because \eqref{eq:alpha_carbons} slightly overpredicts single nutrient contributions to growth in this case (this effect however vanishes when dealing with $E\geq5$ nutrients).

Finally, we compare the biomass production predictions of our \OptimizedSyn~model \eqref{eq:secondorder} against FBA predictions for {\em E. coli}  in media with a fixed number of non-zero random nutrient uptakes normalized to 1 (Methods). 

Figure \ref{fig:exper_beta}a~ shows the \OptimizedSyn~model is able to predict with high accuracy the growth rates computed by using FBA assuming known uptakes.
 The average relative error $\Delta:=\frac{|g_{\rm FBA}-g_{\rm model}|}{g_{\rm FBA}}$ computed over 
500 different random growth media with fixed number of uptakes is systematically smaller for \OptimizedSyn~model predictions than for those of the IM. Notably, the gap between the two models increases with the number of uptakes, due to the more synergistic contributions that are being neglected by the IM model.

Since sugars are the main source of carbons and are quite commonly included in experimental growth media, to reproduce these media we always allow the uptake of one sugar. For more random nutrient setups we find $\Delta$ of the \OptimizedSyn~to be slightly larger, but still consistently smaller than the IM theory (see Appendix). 

\subsection{\label{subsec:ExpComp}Comparison with experiments}

After validating our model {\em in silico}, we test here how well the OS model predicts
 actual growth rates {\em in vivo}. To do so, we compare our model with experimental measurements of
nutrient  uptakes and growth for bacterial culture on complex media. 
 Note that obtaining such type of data is generally not straightforward as measurement of multiple
 uptakes is typically hard. Additionally,  to date, standard experiments used to validate FBA generally
 focus on the simpler case of growth media with a single source of carbon. Nevertheless, a very interesting study
  on complex media where bacterial growth rate and variation of nutrient concentration are measured
  was published by Beg {\em et al.} \cite{beg07}. The authors performed there some {\em E. coli}
 batch culture experiments that allowed them to estimate those quantity simultaneously as a function of time. 
 From their published data, we were able to recover the nutrient uptakes corresponding to every measured
 growth rate (Appendix) and to use such uptakes as inputs in our model.
 This approach allowed us in turn to compare the predicted growth rate with the experimental one.

The results are reported in Fig. \ref{fig:ExpComp}, where we compare  OS model predictions
 with the experimentally measured growth rates.
 Note that now that physiological uptake and growth values are measured,
 we can use proper ${\rm mmol~gDW^{-1}h^{-1}}$ units for the former and $h^{-1}$ for the latter.
 When doing so, model \eqref{eq:secondorder} reaches a remarkable accuracy, especially taking into account that {\em i.)} the {\em E. coli} strain in the
 experiments differs from the reconstruction at our disposal and {\em ii.)} we used the $\boldsymbol{b}$ and $\boldsymbol{a}$ parameters 
 we derived by calibrating the model with FBA, rather than estimating them {\em ad hoc},
 thus  highlighting the broad applicability of our model.

 The excellent agreement we found between the growth predicted
 by our model and the actual growth on a complex medium supports that scaling and synergy
 really are two principles regulating microbial growth {\em in vivo} besides their role in modeling
 metabolism {\em in silico}.

\section{\label{sec:scope}Discussion: Scope and potential limitations of our approach} 
We have used FBA predictions under growth optimization as a reliable source of growth rates, that is, as a substitute for growth experiments with real bacteria. Thus, even though our model is ultimately independent of FBA (in that Eq.~(\ref{eq:secondorder}) does not rely in any way on FBA or on any particular metabolic reconstruction), one may argue that our model is susceptible to suffer the shortcomings of FBA. Here we discuss these shortcomings, although the comparison to experimental data in Fig.~\ref{fig:ExpComp} demonstrates that, whatever limitations FBA may have, our model is able to reproduce experimental growth rates in a variety of realistic conditions.

The first issue is the determination of the so-called ATP maintenance flux. 
This is an additional reaction flux that FBA adds to the set of metabolic reactions and constraints 
to reproduce the experimental growth rates.
Such ATP flux encompasses a series of external factors that 
affect microbial growth rates, such as the uptake rate of nutrients, oxygen availability, and regulation or temperature. 
But although ATP maintenance rates obtained for a specific minimal medium 
have been shown to reproduce accurate results in different growth conditions for certain organisms \cite{feist07}, 
it cannot be assumed that specific values are valid to make predictions for different growth conditions in general. 
To overcome this, we proceed as in \cite{Seaver2012} and first 
evaluate the ATP needed for the polymerization of biomass components by using the values experimentally determined 
(which are available in the literature \cite{feist07,neidhardt90}) and then fix the ATP maintenance to this baseline, 
removing any further ATP maintenance contribution. 
In any case, it is always possible to rescale our findings a posteriori in the same way ATP maintenance is fitted within the FBA approach. Moreover, Fig.~\ref{fig:ExpComp} suggests that the effect of the maintenance flux is not very relevant.

Another caveat of FBA is that it systematically predicts the simultaneous uptake of different sugars, while it is known that microbes absorb their preferred sugar first \cite{monod66}. For this reason FBA will regularly over-predict biomass production in presence of multiple sugars \cite{hong96}. In our approach this is mostly irrelevant because we are concerned with determining growth {\em given} the uptakes of nutrients. In any event, to avoid validating our model against unrealistic settings, we focus on complex growth media containing a single sugar (Methods and Appendix).

Finally, it has been empirically demonstrated that under certain conditions, unicellular organisms do not strictly follow a maximal growth principle \cite{bordel13}. However, it has also been shown that in many occasions the metabolic state predicted by growth maximization is very similar to that of the maximization of other functions \cite{schuetz12}, so that our formalism could be applicable to these conditions.

\section{\label{sec:conclusions}Conclusions}
In this work, we present a second order phenomenological model of metabolism that, by relying on a very limited set of parameters, is able to predict the biomass production of {\em E. coli} in arbitrary complex growth media within 1\% of the actual value for growth {\em in silico} and with great accuracy for growth {\em in vivo}.

Our model shows that  nutrients within the same class are {\it effectively} catabolized in a similar manner, so that the contribution to growth  in the presence of a given nutrient is fully determined by the nutrient's effective carbon content and the class it belongs to. We find that the synergy developed by the uptake of several nutrients increases the catabolic potential of the metabolic network. Such  synergy between nutrients pairs depends on the relative abundance of the nutrients and is capped by the less abundant nutrient.

Our model shows that, effectively, nutrient contributions to growth can be well approximated by the sum of the independent contribution of each nutrient and a synergy contribution. The synergy contribution depends exclusively on nutrient pair synergies so that uptake fluxes are allocated among pair synergies in order to maximize the synergy contribution with the available resources. In this way, the function maximization principle (usually growth) that determines the metabolic state of a unicellular organism can be effectively understood as the optimization of nutrient synergies.


\section*{ Methods}
{\small
\noindent
{\bf Random flux uptakes generation} \\
For each fixed number of uptakes $E$, we generate a vector $\boldsymbol{\phi}$ of uptake fluxes that allows the bacterium to catabolize a combination of fatty acids, 
amino acids and bases, plus one sugar only. To do so, only one of the entries of $\boldsymbol{\phi}$ that do correspond to sugar uptakes is chosen uniformly at random to have a value different from zero. Such value is uniformly drawn at random in the range $(0,1) {\rm~arb. units}$. All $E-1$ remaining uptakes are uniformly chosen at random among entries of $\boldsymbol{\phi}$ that do not correspond to a sugar. Again, the flux value is drawn in the range $(0,1) {\rm~arb. units}$. After all the $E$ nonzero entries of $\boldsymbol{\phi}$ are drawn, we normalize the uptakes so that the total uptake is always equal to one (see Appendix for results in other complex media).

\noindent
{\bf Optimal synergy model}

Suppose we want to compute the growth of a vector $\mathbf{\phi}$ of uptake fluxes with $E$ non-zero entries according to the \OptimizedSyn~model \eqref{eq:secondorder}.

In order to allocate the uptake of fluxes to maximize synergy we proceed as follows.
First, we compute all $E(E-1)/2$ synergies $\beta'$ and rank them according to their corresponding contributions to growth from largest to smallest. Starting from the largest, we evaluate which nutrient in the pair $(n_1,n_2)$ is in excess by comparing the flux ratio $C_{n_1} \phi_{n_1}/(C_{n_2}\phi_{n_2})$ to the transition value $T(n_1,n_2) = b_{n_2 n_1}/b_{n_1 n_2}$ of the corresponding $\beta'$ function. For instance, if $C_{n_1}\phi_{n_1}/(C_{n_2}\phi_{n_2})<T(n_1,n_2)$, $n_2$ is in excess. We then store this contribution, set the limiting flux $\phi_{n_1}$ to zero and  reduce $\phi_{n_2}$  by its distance from the transition value as $\phi_{n_2} \to \phi_{n_2} - C_{n_1}/C_{n_2}\phi_{n_1} T(n_1, n_2)$. Note that this implies that $\phi_{n_1}$ is not used in other synergies. All the other fluxes are kept constant. These updated fluxes are used to re-compute the synergies occupying lower positions in the rank, and the process is repeated for the second largest $\beta'$. In this way synergies at position $k$ in the rank are computed with effective fluxes $(\phi_{n_1}^k,\phi_{n_2}^k)$ that take into account both the limitedness of resources and their optimal routing. 

A slightly different version of our approach, where ranking of synergies is computed after each step $\phi_n^k \to \phi_n^{k+1}$ is not as accurate as the protocol described above (see Appendix and fig. \ref{fig:exper_beta}).

}
\begin{figure*}[t!]
\centering

\includegraphics[width=\textwidth]{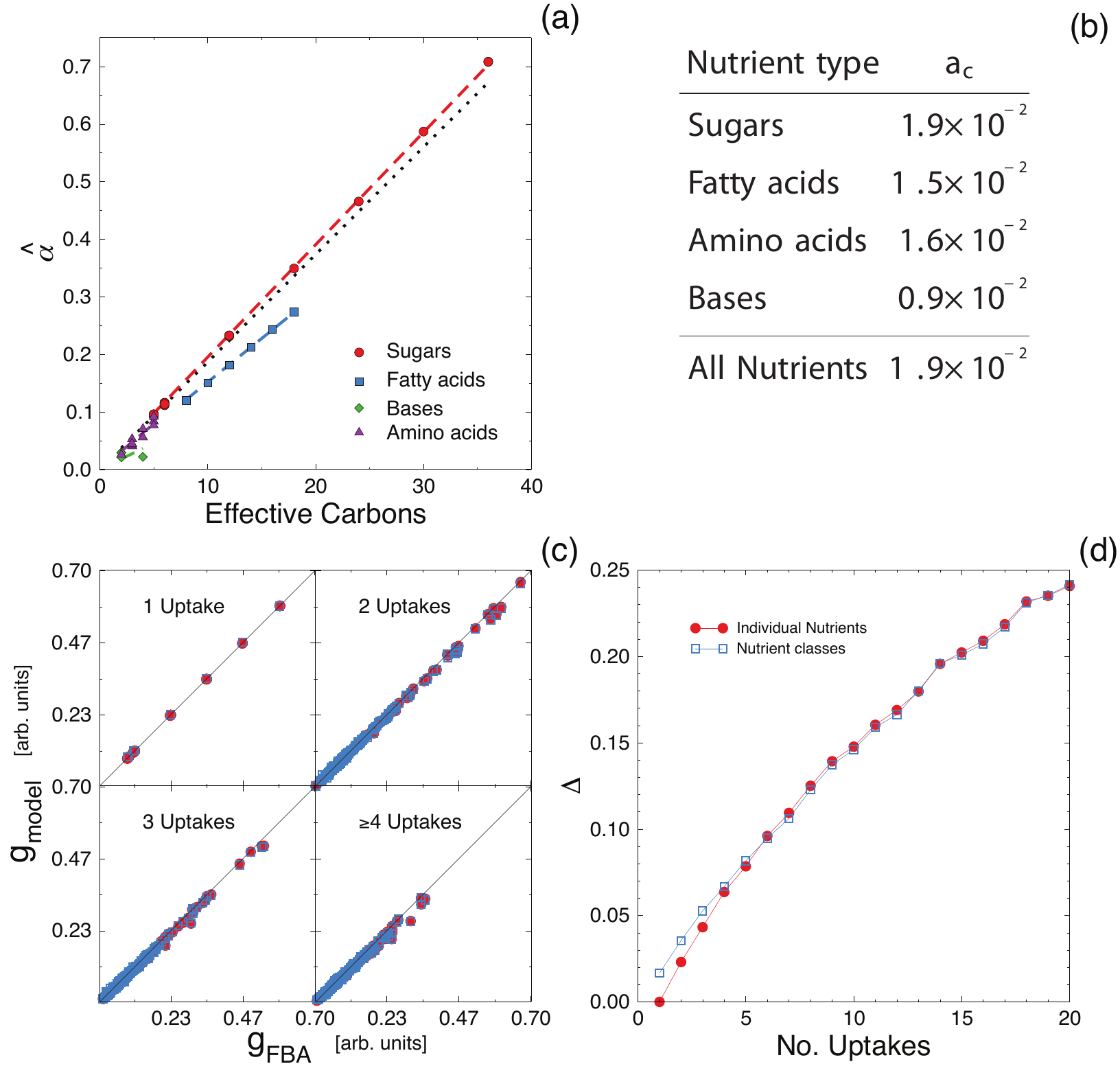}
\caption{Idealized metabolism theory. {\bf (a)} The $\hat{\alpha}$ parameters introduced in \eqref{eq:f_1_order}, versus the number of effective carbons for each of the nutrients 
considered in our study. We consider  nutrients  in four groups: sugars, fatty acids, bases and amino acids. 
The $\hat{\alpha}$ coefficients are a linear function of the effective number of carbons whose slope depends very weakly on the nutrient class, except for bases (see panel b). The dashed lines 
show linear fits for each class of nutrients, while the black dotted line is a fit considering all of them together.
{\bf (b)} The coefficients $a_c$ introduced in \eqref{eq:alpha_carbons}. We show the values of $a_c$ obtained from the fits shown in panel a).
$a_c$ varies weakly with nutrient class.
{\bf (c)} Predictions of the idealized metabolism theory, \eqref{eq:f_1_order}, versus FBA results for a selection of 100 random media with increasing number of possible uptakes (see Methods). Filled red circles correspond
 to using exact $\alpha$ values, while empty blue squares to \eqref{eq:alpha_carbons}.
{\bf (d)} The relative error $\Delta = \frac{|g_{\rm FBA}-g_{\rm model}|}{g_{\rm FBA}}$ of the IM theory predictions 
for the two different choices of $\hat{\alpha}$ averaged over 500 random media, for increasing number of uptakes. 
$\Delta$ is relatively small in presence of a few nutrients only, but it increases roughly linearly. Note that the error performed
when using \eqref{eq:alpha_carbons} in presence of one nutrient only is different from zero, meaning that \eqref{eq:alpha_carbons}
 does not correctly capture single nutrient contributions to growth. This effect however is negligible increasing the number of
  nutrients, as the two $\Delta$ curves overlap.
}
\label{fig:alphas}
\end{figure*}
\clearpage

\begin{figure*}[t!]
\includegraphics[width=\textwidth]{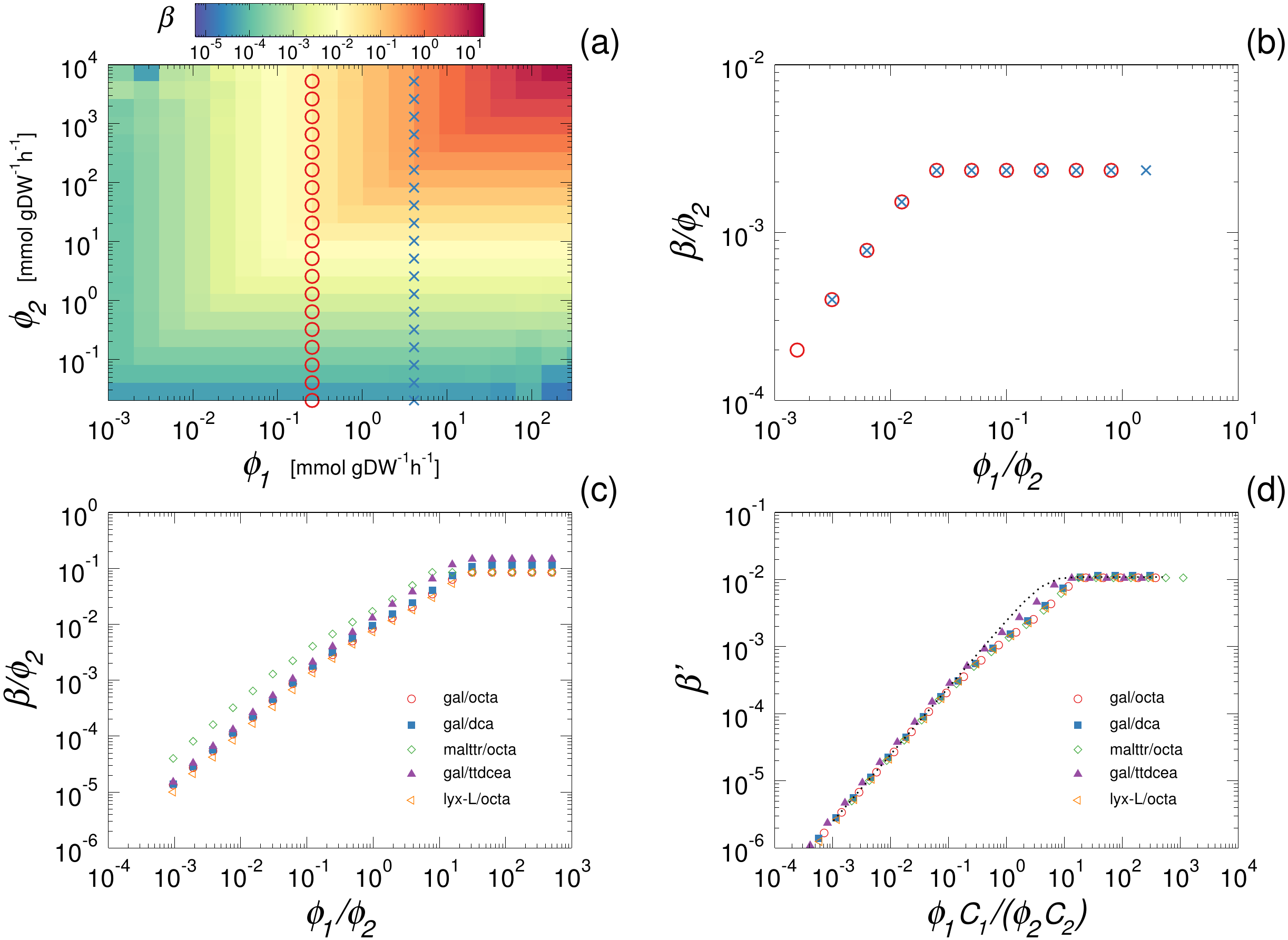}
\caption{Scaling of nutrient synergy contributions. 
{\bf (a)} The function $\beta$, \eqref{eq:beta}, that expresses the gap between the linear model predictions \eqref{eq:f_1_order} and the FBA 
results for the growth rate of {\em E. coli}, when there are two nutrient uptakes different from zero. 
We show here the simultaneous uptake of dodecanoate and butyrate (both fatty acids) as a typical example. 
$\beta$ is a growing function of the exchange fluxes of both nutrients. 
The circles and crosses correspond to the two (example) curves that are shown, once rescaled, in panel (b).
{\bf (b)} Scaling property of $\beta$, \eqref{eq:scale_beta}. We plot the same data points of panel (a): each curve shows $\beta/\phi_2$ as a function of $\phi_1/\phi_2$, for two different fixed values of $\phi_1$. Such normalization allows to collapse all points on the same curve.
{\bf (c)} The function \eqref{eq:scale_beta} for a set of five sugar-fatty acid pairs,  that shows a characteristic linear--plateau behavior. 
{\bf (d)} The rescaling property \eqref{eq:betaCarbons}. We rescale the uptake fluxes of the nutrient pairs shown in panel c with the number of carbons of each nutrient. All the points collapse on the same curve. The dotted line corresponds
to the function \eqref{eq:tanhbeta}, 
where we set $\overline{b}_{_{\rm s\cdot fa}}, \overline{b}_{_{\rm fa\cdot s}}$ as the average of the set $\boldsymbol{b}_{_{\rm s\cdot fa}}, \boldsymbol{b}_{_{\rm fa\cdot s}}$ for all the sugar--fatty acid pairs.}
\label{3dbeta}
\end{figure*}

\begin{figure*}[t!]
\begin{center}
\includegraphics[width=0.9\textwidth]{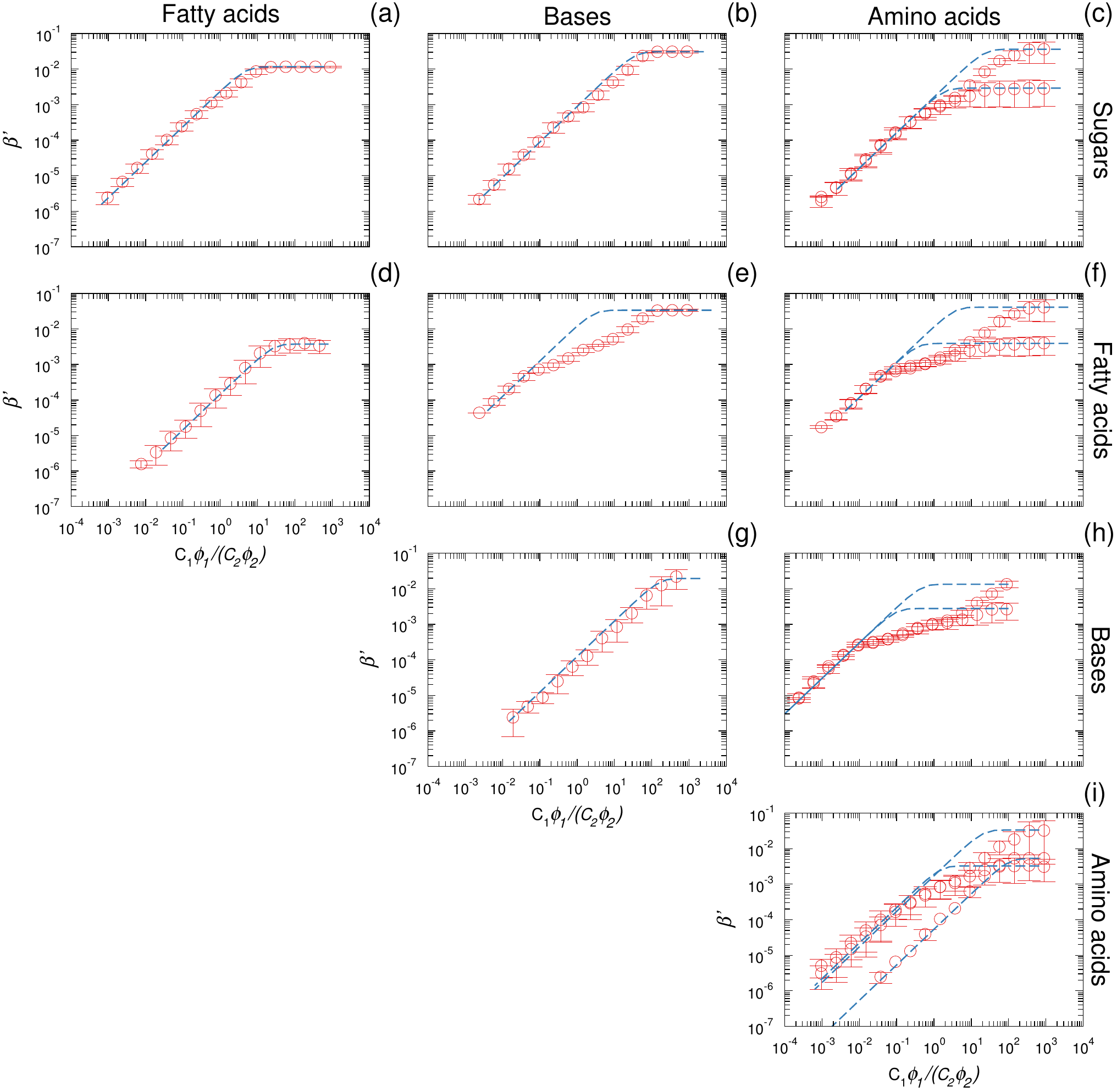}
\end{center}
\caption{Nutrient synergy contributions.
We show the $\beta'$ function, \eqref{eq:betaCarbons}, for pairs of four nutrient classes: sugars, fatty acids, bases and amino acids. 
Dashed lines correspond to the function in \eqref{eq:tanhbeta} where the parameters $\{b_{\kappa\jmath}\}$ are averaged over all pair of nutrients in the corresponding pair of classes.
}
\label{fig:all_betas}
\end{figure*}
\begin{table*}[t!]
\caption{Average numerical values of the parameters of the phenomenological model in \eqref{eq:tanhbeta}. We show here the average slope ($b_{12}$) and plateau ($b_{21}$) values of the $\beta'$ functions for the cross interactions plotted in Fig. \ref{fig:all_betas}. For nutrient pairs involving an amino acid we obtain two different plateau values, depending on the metabolic processes in which the amino acid participates (see text). If two amino acids are involved, also an additional slope is needed. When the pair is inverted, for different nutrient classes, the values of the plateau and and slope are also swapped. Note that we order nutrients according to their carbon content and do not consider pair permutations. For this reason, for pairs of the same class (e.g. Fatty acids--Fatty acids), values $\overline{b}_{12}$ and $\overline{b}_{12}$ are not equal: $\overline{b}_{12}$ captures growth on media where the nutrient with more carbons is in excess, while $\overline{b}_{12}$ renders the opposite situation.}
\begin{tabular}{|l|m{4ex}|c|c|c|c|} \hline \hline
&
\setlength{\unitlength}{4ex}
\begin{picture}(1,1)
\put(0.1,-0.2){$1$}
\put(0.7,0.5){$2$}
\end{picture}
&  Fatty acids  & \phantom{spce}Bases\phantom{spce} & \multicolumn{2}{c|} {Amino acids}\\ \hline\hline

\multirow{2}{*}{ Sugars }& $\overline{b}_{12}$ & \parbox{0.2\textwidth}{\centering $2.4 \times 10^{-3}$ } & $8.8 \times 10^{-4}$ & \multicolumn{2}{c|} {$1.6 \times 10^{-3}$} \\
\cline{2-6}
&$\overline{b}_{21}$ & $1.2 \times 10^{-2}$ &  $3.1 \times 10^{-2}$ & $2.9 \times 10^{-3}$ & $3.6 \times 10^{-2}$ \\ \hline

\multirow{2}{*}{ Fatty acids }& $\overline{b}_{12}$ & $1.4 \times 10^{-4}$ &  $1.2 \times 10^{-2}$ & \multicolumn{2}{c|}{$1.2 \times 10^{-2}$} \\
\cline{2-6}

&$\overline{b}_{21}$ & $3.5 \times 10^{-3}$ & $3.4 \times 10^{-2}$ & $3.9 \times 10^{-3}$ & $4.1 \times 10^{-2}$ \\ \hline

\multirow{2}{*}{ Bases }& $\overline{b}_{12}$ & & $7.2 \times 10^{-4}$  & \multicolumn{2}{c|}{$3.0 \times 10^{-2}$ } \\
\cline{2-6}

&$\overline{b}_{21}$ & & $1.3 \times 10^{-2}$ & $2.8 \times 10^{-3}$ & $1.3 \times 10^{-2}$ \\ \hline

\multirow{2}{*}{ Amino acids }& $\overline{b}_{12}$ & & & $2. \times 10^{-3}$ & $5.4 \times 10^{-5}$ \\
\cline{2-6}

&$\overline{b}_{21}$ & & & $4. \times 10^{-3}$ & $3.3 \times 10^{-2}$ \\ \hline
\end{tabular}
\label{tab:summary}
\end{table*}
\begin{figure*}[t!]
\begin{center}
\includegraphics[width=1.\textwidth]{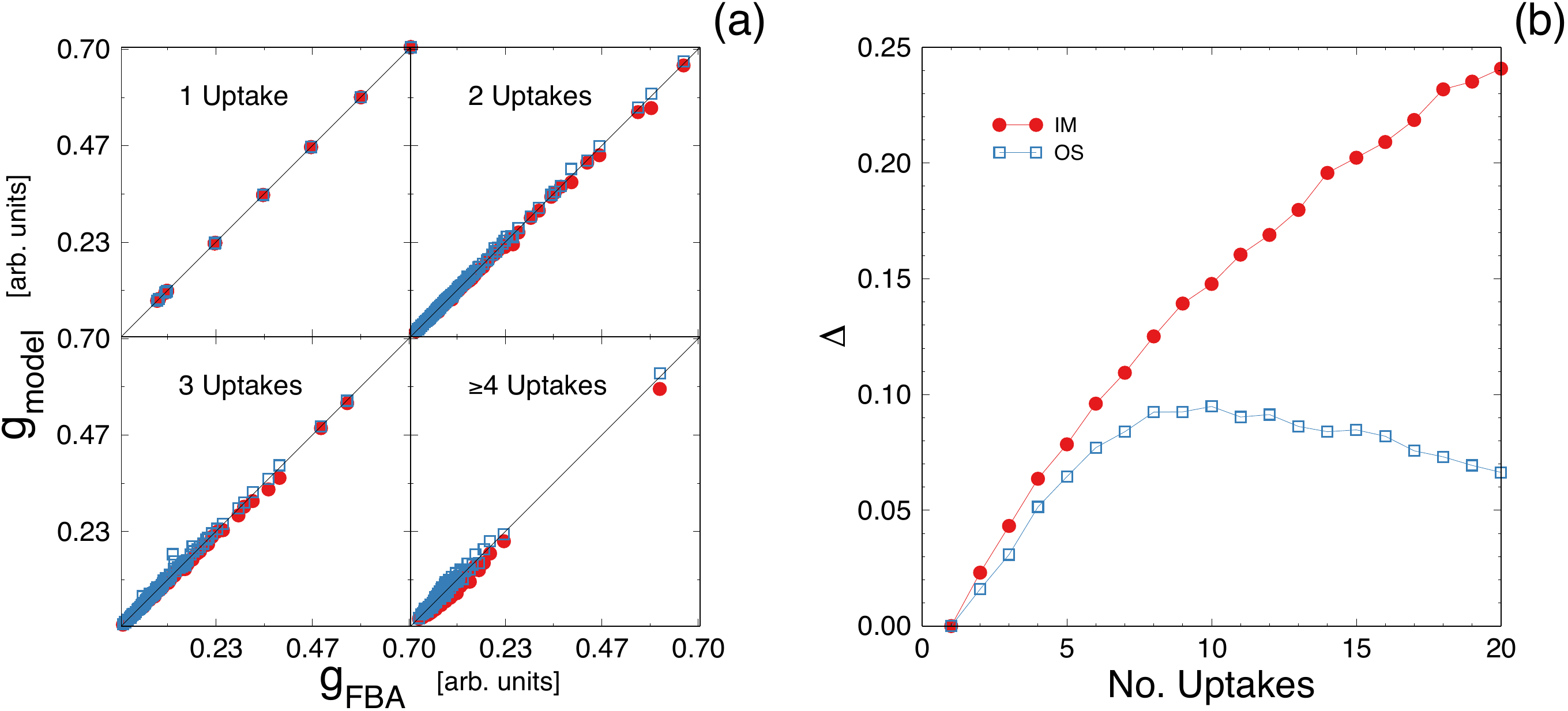}
\end{center}
\caption{Second order equitative synergy theory. 
{\bf (a)} Predictions of the optimized synergy model (\OptimizedSyn) \eqref{eq:secondorder}, empty blue squares, versus the FBA results, compared with the IM theory \eqref{eq:f_1_order}, filled red circles, for 100 different random media at increasing number of uptakes (see Methods and Appendix for the details on growth media). Here, we use the exact values of parameter $\hat{\alpha}$ and the average interclass value of parameters $b$. 
{\bf (b)} The relative error $\Delta=\frac{|g_{\rm model}-g_{\rm FBA}|}{g_{\rm FBA}}$ vs. the number of uptakes for the IM (filled red circles) and the \OptimizedSyn~model (empty blue squares), averaged over 500 different random media.
The relative error of the IM theory grows almost linearly, while it remains much lower in the \OptimizedSyn~model and becomes roughly independent of the number of uptakes for $E\geq 6$.}
\label{fig:exper_beta}
\end{figure*}
\begin{figure*}
\centering
\includegraphics[width=0.5\textwidth]{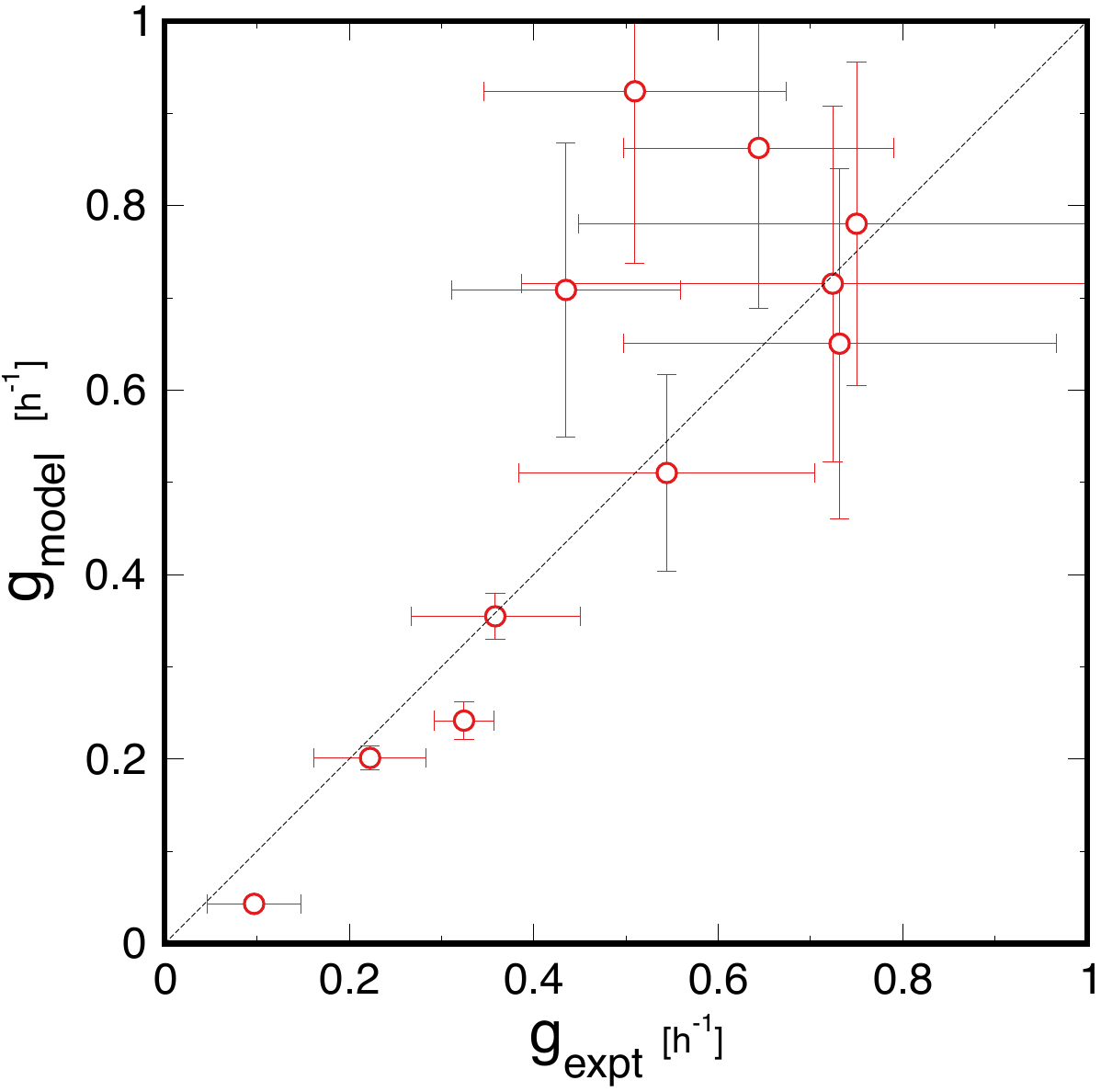}
\caption{Comparison of the OS model, \eqref{eq:secondorder}, (y axis) with the experimental growth of Beg {\em et al.}, \cite{beg07} (x axis); the dashed diagonal indicates perfect agreement. The uptakes corresponding to each experimental growth rate were computed (Appendix) and used as an input of the OS model to evaluate the predicted growth. The x error bars are one standard error, the y error bars indicate all feasible growths consistent with the uptakes plus/minus their error. We find a fair agreement between our theory and the experimental measurements, supporting that scaling and synergy are two principles regulating also microbial growth {\em in vivo}.
}
\label{fig:ExpComp}
\end{figure*}
\begin{table*}[t!]
\caption{The metabolic pathways included in the logistic model to predict amino acids groups ($\HighSyn$~or $\SmallSyn$). We report in the first column the pathway names, sorted for decreasing Bayesian Information Criterion associated with the model. In the second column we list the number of amino acids participating in each pathway.}
\begin{tabular}{|c l|c|}
\hline
\hline
&Metabolic pathway & No. a. acids\\
\hline
\hline
{\small 1.}& alanine, aspartate and glutamate metabolism& 6\\
{\small 2.}& valine, leucine and isoleucine degradation& 2\\
{\small 3.}& phenylalanine, tyrosine and tryptophan biosynthesis& 3\\
{\small 4.}& sulfur relay system& 2\\
{\small 5.}& glycine, serine and threonine metabolism& 7\\
{\small 6.}& arginine and proline metabolism & 7\\
\hline
\end{tabular}
\label{tab:Aacids}
\end{table*}
\begin{acknowledgments}
This work was supported by a James S. McDonnell Foundation Research Award, Spanish Ministerio de Econom\'ia y Comptetitividad (MINECO) Grant FIS2013-47532-C3, European Union Grant PIRG-GA-2010-277166, European Union Grant PIRG-GA-2010-268342, and European Union FET Grant 317532 (MULTIPLEX)
LANA acknowledges the support of NSF award SBE 0624318 Foundation and the W.M. Keck Foundation.
\end{acknowledgments}

\clearpage
\appendix

\section{The metabolic reconstruction}\label{sec:reconstruction}
We use the genome scale {\em E. coli} metabolic reconstruction iAF1260 \cite{feist07}. Such reconstruction features 1678 metabolites and 2392 reactions, of which 299 are exchange reactions. The minimal medium is composed by 18 essential nutrients Ca2, cobalt2, Cu2, Zn2, Mn2, cbl1, H$_2$O, Pi, H, K, Cl, Fe2, Fe3, mobd, Na1, Nh4, So4, Mg2 \cite{feist07}. The fluxes of the reactions that uptake these nutrients are always kept different from zero. In our analysis we assume nutrient uptakes are known. Thus we focus exclusively on the 63 exchange reactions delivering sugars (22 reactions), fatty acids (6 reactions), amino acids (26 reactions), and bases (9 reactions) to the bacterium (see Table \ref{tab:allNutes}), and keep all other exchanges locked to zero.

\section{Flux Balance Analysis}\label{sec:fba}
Flux Balance Analysis (FBA) is a mathematical tool to predict, under certain assumptions, the fluxes $\boldsymbol{\nu}$ and the biomass production $g_{\rm FBA}$ of a metabolic network \cite{Orth2010}. Given the stoichiometry $\boldsymbol{S}$ of the network, FBA aims at finding the solution of the metabolic mass balance equation under steady state condition. Denoting by $\boldsymbol{c}$ the vector of metabolic concentration, FBA seeks thus to solve the system of linear equations:
\begin{equation}\label{eq:fba}
\dot{\boldsymbol{c}} = \boldsymbol{S}\boldsymbol{\nu}=0.
\end{equation}
Since in real metabolic networks there are much more reactions than metabolites, the above system is underdetermined and it allows several solutions. From the space of solutions, physiologically relevant points are usually selected by coupling the mass balance problem \eqref{eq:fba} with an optmization principle. Quite generally, thus, a FBA problem seeks solutions to \eqref{eq:fba} such that a linear objective function $Z$ of the form
\begin{equation}\label{eq:fbaObj}
Z = \sum_kr_k\nu_k,
\end{equation}
with $r_k$ some positive constants, is maximized. The objective function is often related to the biomass production.
~In our case we focus solely on the maximization of biomass polymerization, so that we have one flux only appearing in the sum \eqref{eq:fbaObj} (which expresses the biomass synthesis) and we can assume $Z=g_{\rm FBA}$. Finally, we note that when essential nutrients are assumed to available in excess, \eqref{eq:fba} specifies a linear problem that is defined up to multiplicative constant: any solution to \eqref{eq:fba} may be rescaled through a constant factor and still be a valid solution. We therefore keep uptakes in arbitrary units when validating our model against FBA.
\begin{table*}[t!]
\caption{The 63 uptake fluxes considered in our study. We include uptakes delivering sugars (22 reactions), fatty acids (6 reactions), amino acids (26 reactions), and bases (9 reactions) to the bacterium.}

{\footnotesize %
\begin{tabular}{|rlrl|rl|rlrl|rl|}
\hline
\multicolumn{4}{|c|}{Sugars}&\multicolumn{2}{c|}{Fatty acids}&\multicolumn{4}{c|}{Amino acids}&\multicolumn{2}{c|}{Bases}\\
\hline
1. & L-Arabinose &14. & Maltose &1. & Octanoate&1. & Glycine&14. & D-Methionine &1. & Allantoate\\
2. & L-Lyxose &15. & Melibiose &2. & Decanoate&2. & D-Alanine &15. & L-Methionine&2. & Cytosine\\
3. & D-Ribose &16. & Sucrose &3. & Dodecanoate &3. & L-Alanine &16. & Ornithine &3. & Uracil\\
4. & D-Xylose &17. & Trehalose &4. & Tetradecanoate&4. & D-Cysteine &17. & L-Proline &4. & Adenine\\
5. & L-Xylulose &18. & Maltotriose &5. & Hexadecanoate&5. & L-Cysteine &18. & L-Valine &5. & Guanine\\
6. & D-Allose &19. & Maltotetraose &6. & Octadecanoate&6. & D-Serine &19. & L-Arginine &6. & Hypoxanthine\\
7. & D-Fructose &20. & Maltopentaose &&&7. & L-Serine &20. & L-Histidine &7. & Orotate\\
8. & L-Fucose &21. & 1-4-$\alpha$-D-glucan&&&8. & L-Asparagine &21. & L-Isoleucine &8. & Thymine\\
9. & $\beta$-D-Galactose &22. & Maltohexaose&&&9. & L-Aspartate &22. & L-Leucine&9. & Xanthine\\
10. & Galactose &&&&&10. & L-Homoserine &23. & L-Lysine &&\\
11. & D-Mannose &&&&&11. & L-Threonine &24. & L-Phenylalanine &&\\
12. & L-Rhamnose &&&&&12. & L-Glutamine &25. & L-Tyrosine &&\\
13. & Lactose &&&&&13. & L-Glutamate &26. & L-Tryptophan &&\\
\hline
\end{tabular}
}
\label{tab:allNutes}
\end{table*}
\section{Generation of the growth media}\label{sec:rmedia}
We focus only on nutrients that can be uptaken by the organism and produce growth \cite{Seaver2012}. The growth media we generate therefore only contain sugars, fatty acids, amino acids, and bases. Since multiple uptake of sugars is not observed \cite{monod66}, we allow for the exchange of one sugar only and randomly allow all other nutrients to be uptaken by the bacterium. Summing up all the exchange fluxes listed in Sec.\ref{sec:reconstruction}, each growth medium can therefore be composed of 42 nutrients at the most (i.e. one sugar and 41 other nutrients), plus the 18 nutrients in the minimal medium.

As the minimal medium is always included, just considering the 22 sugars and the 41 remaining nutrients, for each growth medium we hence have a 63--dimensional random vector of exchange fluxes $\boldsymbol{\phi}$ which, for any fixed number of uptakes $E$, is generated as follows (see Fig. \ref{fig:randomMedia} for a pictorial representation of the growth media):
\begin{figure*}[t]
\centering
\includegraphics[width=0.9\textwidth]{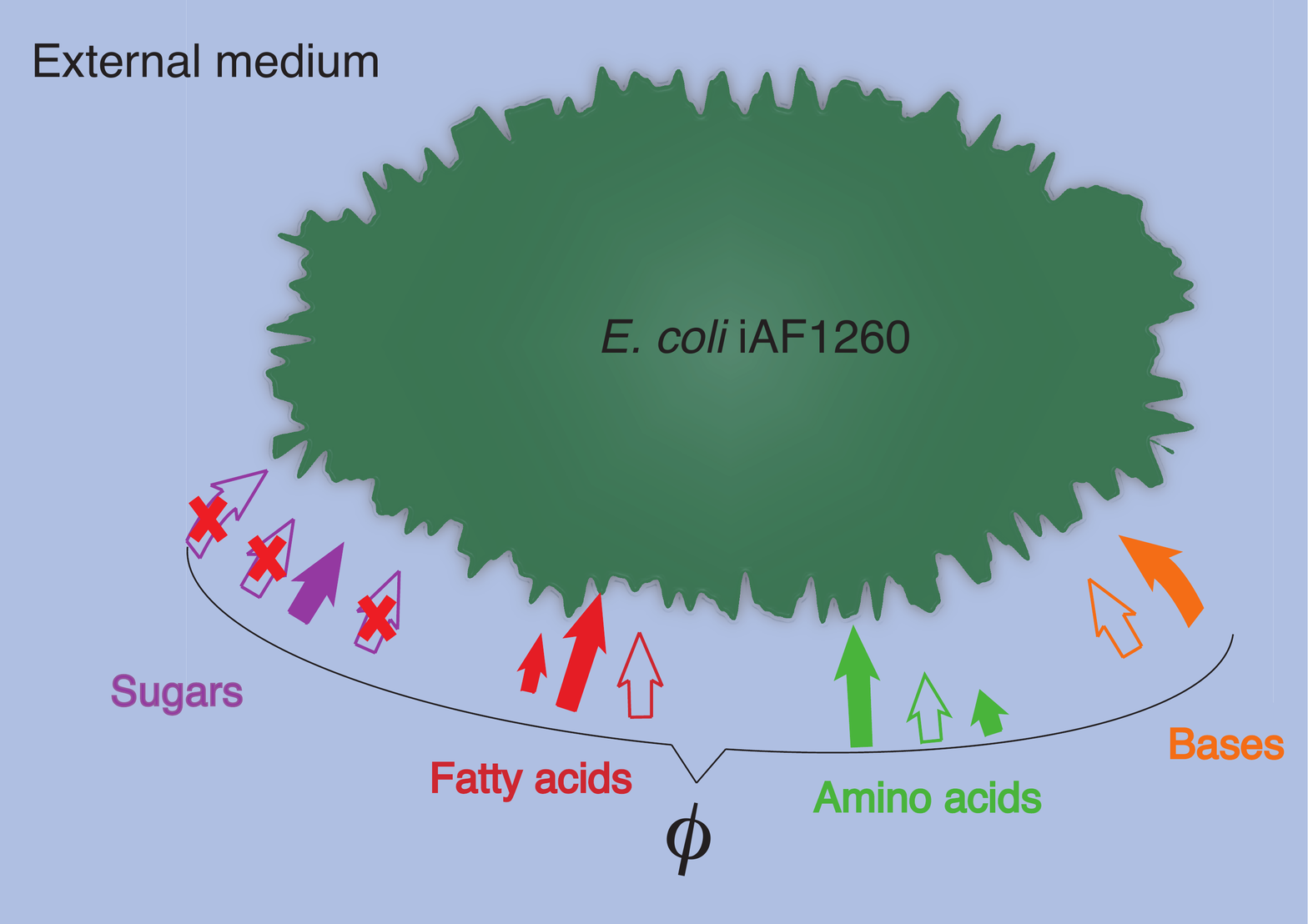}
\caption{Illustration of how random media are generated. Besides the minimal medium, we only consider growth on sugars, fatty acids, amino acids, and bases. Each random medium we generate only contains one sugar (the purple filled arrow), plus a set of other nutrients. The sugar and the remaining nutrients are all uniformly chosen at random. These nutrients and their uptake value form a random vector of exchange fluxes $\boldsymbol{\phi}$. In the figure we sketch as filled arrows all the nutrients included in the random medium and as empty arrows the ones not considered. For any random medium considered, uptakes are normalized so that $\sum_i\phi_i = 1~{\rm arb. units.}$.}
\label{fig:randomMedia}
\end{figure*}
\begin{itemize}
\item Only one of the 22 entries delivering sugars is uniformly chosen at random. We randomly fix its value uniformly in the set $\phi_{\rm sug}\in(0.,1.)~{\rm~arb. units}$.
\item The remaining $E-1$ uptakes are uniformly drawn at random among the 41 entries of $\boldsymbol{\phi}$ that do not correspond to a sugar. The value of each flux is again uniformly drawn at random in the set $(0.,1.)~{\rm~arb. units}$.
\item The $E$ nonzero entries of $\boldsymbol{\phi}$ are normalized so that $\sum_i\phi_i=1~{\rm~arb. units}$
\end{itemize}
In all the complex growth media we generate we always include the essential nutrients, which are assumed to be present in excess, {\em i.e.} they are uptaken at a rate $1\times10^{7}~{\rm arb. units}$, equivalent to infinite uptake rate in the metabolic reconstruction.
\section{Selection of the minimal model for the growth on amino acids}\label{sec:BIC}
When studying nutrient--class--wide pairwise interactions involving amino acids, we noticed that the $\beta'$ functions appearing in Fig. \ref{fig:all_betas} tended to acquire two plateau values. We hence divided the amino acids into sets $\HighSyn$~and $\SmallSyn$, according to whether their corresponding $\beta'$ plateau value was above or below $10^{-2}$, respectively.

\begin{figure*}[t]
\centering
\includegraphics[width=0.8\textwidth, angle=-90]{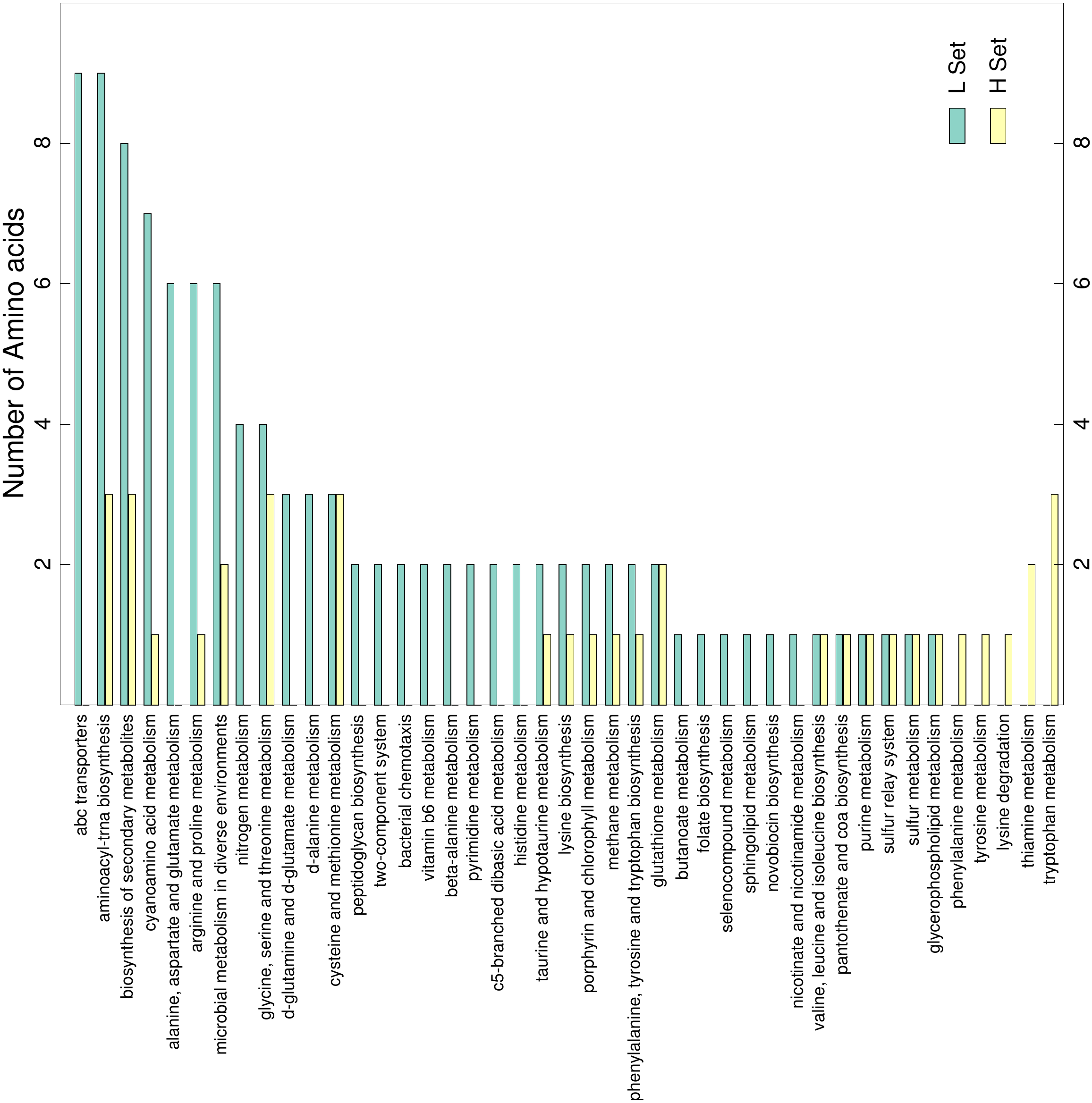}
\caption{Number of amino acids in sets $\HighSyn$~and $\SmallSyn$~for each metabolic pathway. We see that the amount of amino acids in each set is uneven in the majority of pathways, with most of them only featuring amino acids in the $\SmallSyn$~set. We opted to exploit this characteristic to predict to which set each amino acid belongs to and automatically assign it a $\beta'$ plateau value.}
\label{fig:pathsAa}
\end{figure*}

By doing this, we observed that the pathways that process a given amino acid correlate in some way with its associated $\beta'$ plateau values. Indeed, as we show in Fig. \ref{fig:pathsAa}, many metabolic pathways feature either amino acids belonging to only one set, or a far exceeding number of amino acids in one of the two sets.

We thus opted to predict whether a given amino acid belonged to group $\HighSyn$~(or $\SmallSyn$) by exploiting the minimum information on the metabolic processes it participates in. We developed a linear model $\pi_i$ for each amino acid $i$ and used logistic regression to estimate the probability $\mathcal{P}_i(i \in \HighSyn | \pi_i)$ for metabolite $i$ to belong to group $\HighSyn$~given model $\pi_i$. Considering a set $\mathcal{M}$ of $n$ metabolic pathways, we assumed
\begin{equation}\label{eq:linreg}
\begin{split}
\pi_i &\equiv \xi_0+\sum_{j = 1}^n \xi_j X_i^j\\
\mathcal{P}_i(i \in \HighSyn | \pi_i)&=\frac{1}{1+\exp \pi_i},
\end{split}
\end{equation}
where the sum runs over the $n$ pathways in $\mathcal{M}$. In \eqref{eq:linreg} $X_i^j$ is a binary variable taking value 1 if amino acid $i$ participates to pathway $j$ and 0 otherwise. All coefficients $\{\xi_j\}_{j=1}^n$ have real values. For each set $\mathcal{M}$ we estimate $\{\xi_j\}_{j=1}^n$ by miximizing the likelihood $\mathcal{L} = \prod_{i=1,}\mathcal{P}_i$. The coefficient $\xi_0$ is related to the probability that an amino acid $i$ belongs to $\HighSyn$ while not participating to any pathway in $\pi_i$. As we aim to gain the maximum predictive power by exploiting the minimum information, we opted to seek for the smallest set $\mathcal{M}$ that yields the largest rate of correct guesses, that is, which returns $\mathcal{P}_i$ larger than 0.5 for metabolites actually belonging to $\HighSyn$~in the majority of cases. The minimum set may be found by minimizing the Bayesian information criterion (BIC) \cite{schwarz78}
, {\em viz}:
\begin{equation}
\begin{split}
{\rm BIC} &= (n+1)\log N - 2 \log \mathcal{L} 
, 
\end{split}
\end{equation}
where $n \equiv \|\mathcal{M}\|$ is the size of the set $\mathcal{M}$ (i.e. the number of included pathways), $N$ is the number of amino acids and $\mathcal{L}$ is the likelihood that the observed $\HighSyn$, $\SmallSyn$~sets are generated by models $\{\pi_i\}_{i=1}^N$.

\begin{figure*}[t]
\centering
\includegraphics[width=0.8\textwidth]{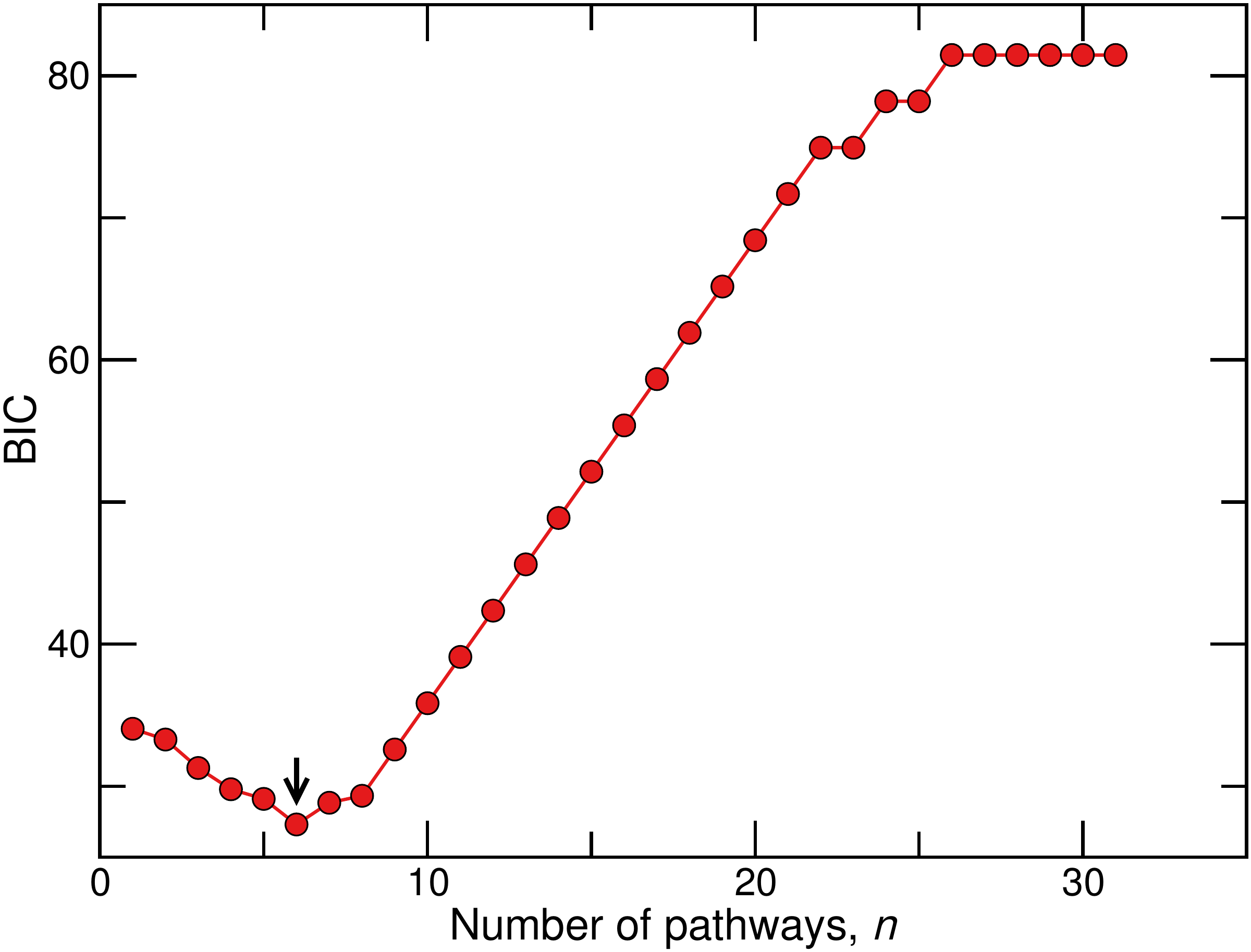}
\caption{The Bayesian Information Criterion as a function of the number of pathways $n$. Starting with zero pathways, we iteratively incorporated into the model \eqref{eq:linreg} the metabolic pathway that yielded the minimum BIC. This allows to gain predictive power and to lower the BIC up to $n=6$ pathways (black arrow). Inclusion of further information does not enhance the predictive ability and only overfits the model.}
\label{fig:bic}
\end{figure*}

To seek for the minimal $\mathcal{M}$, we started out with zero pathways and then used an iterative greedy approach that at each step added the pathway that yielded the minimum BIC, that is, that maximized the likelihood $\mathcal{L}$. The result of this iterative approach is shown in Fig. \ref{fig:bic}: the first point features one metabolic pathway and renders a BIC close to 30. Adding parameters (i.e. adding metabolic pathways) lowers the BIC up to $n=6$ where there is no more significative gain in predictive power and adding more pathways only overfits the model, so that the BIC starts to grow. The whole analysis was performed using R (version 2.15.3 \cite{R05}).

Once we knew the profile of the BIC, we retained the set $\mathcal{M}$ that minimized it. Such set is the best trade off between the likelihood $\mathcal{L}$ (i.e. the predictive power) and the number of pathways included in the model. The six pathways included in the final $\mathcal{M}$ yielded a ${\rm BIC}=27.3$ and are listed in Table \ref{tab:BIC}, where we also report the BIC returned by all models featuring $n\leq6$ pathways and the number of amino acids participating in each pathway included.

\begin{figure*}[t]
\centering
\includegraphics[width=0.7\textwidth]{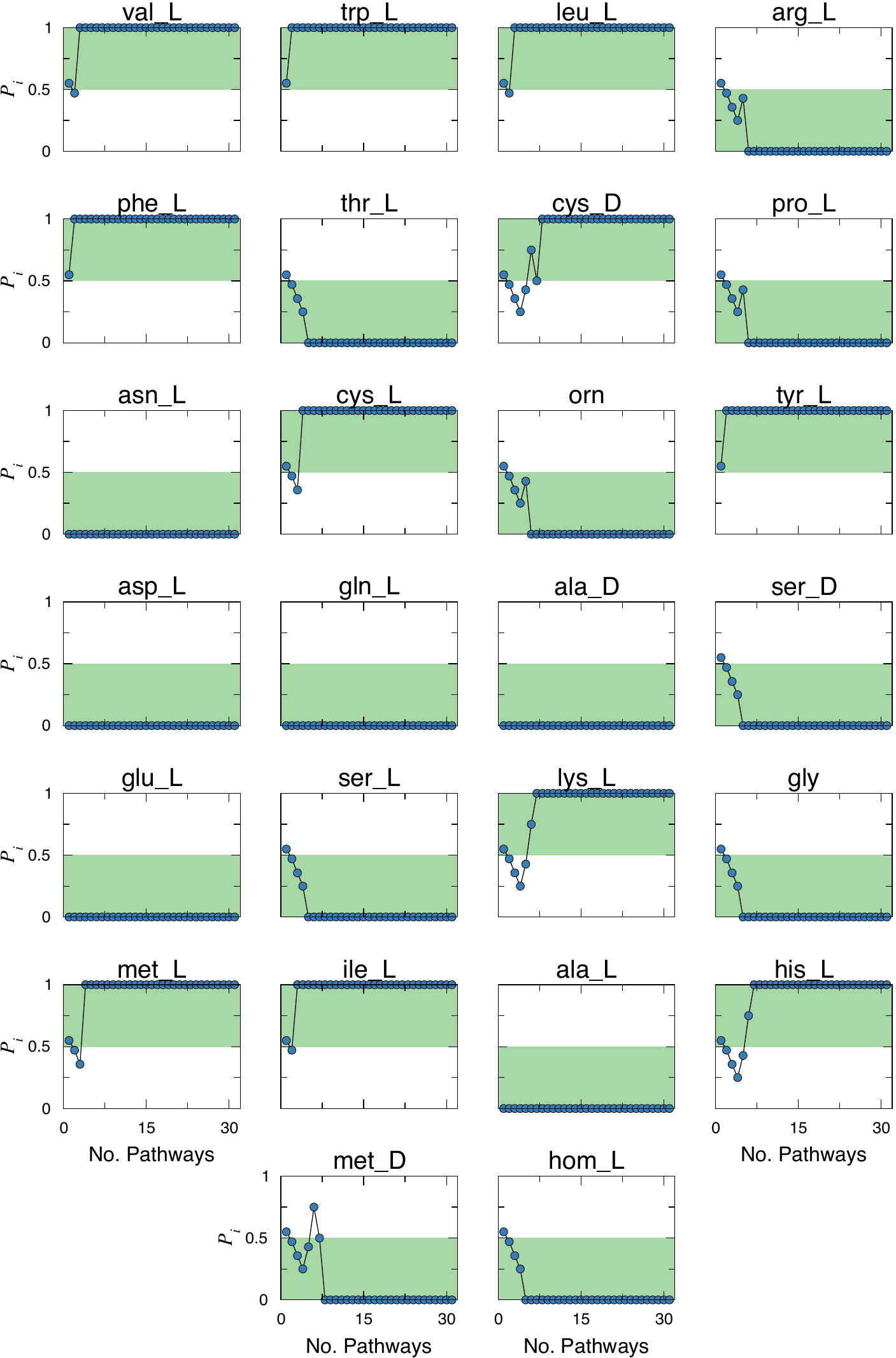}
\caption{The probabilities $\mathcal{P}_i(i \in \HighSyn | \pi_i)$ of each amino acid $i$ varying the number of pathways $n$ included in the model $\pi_i$. The shaded green area highlights the expected region where $\mathcal{P}_i$ should lie, i.e. $\mathcal{P}_i \in [0,0.5]$ and $\mathcal{P}_i \in (0.5, 1]$ for amino acids in sets $\SmallSyn$~and $\HighSyn$~respectively. For the majority of them, the inclusion of only a few pathways in $\pi_i$ is enough to predict the correct set. When $n=6$, that is, when the BIC is minimum, we correctly capture the behavior of all amino acids except for D-Methionine (met\_D).}
\label{fig:Pevolution}
\end{figure*}

In Fig. \ref{fig:Pevolution}, we show the probabilities $\mathcal{P}_i(i \in L | \pi_i)$ as a function of the number of pathways $n$ in the model $\pi_i$. In our analysis we fix a threshold of 0.5 and assume metabolite $i$ belongs to $\HighSyn$~if $\mathcal{P}_i>0.5$ and $i\in S$ otherwise. The green shaded area in Fig. \ref{fig:Pevolution} indicates the region where we expect $\mathcal{P}_i$ to lie: for the vast majority of the amino acids only a few parameters in the $\pi_i$ are sufficient to classify all amino acids into sets $\SmallSyn$ or $\HighSyn$. For the case $n=6$ pathways, which minimizes the BIC, we see that there is only one amino acid which is not correctly classified, namely D-Methionine (met\_D). All the rest of the amino acids are correctly assigned to either $\SmallSyn$~or $\HighSyn$~by only inspecting whether they participate in the metabolic pathways listed in Table \ref{tab:BIC}.

Since knowing whether a given amino acid participates to these six pathways is sufficient to know where its associated $\beta'$ plateau will lie, we decided to model the $\beta'$ functions through their phenomenological form \eqref{eq:betaPar} and assign two possible values to parameters $\boldsymbol{b}$, which are evaluated by averaging $\beta'$ corresponding to amino acids in the sets $\HighSyn$~and $\SmallSyn$~separately. 

\begin{table*}[t!]
\caption{The six pathways included in the model $\pi$ that minimizes the Bayesian Information criterion. We report in each row the name of the pathway, the number of amino acids participating in it, and the BIC value of the model containing all pathways up to the row, so that the last line has the minimum BIC value.}
\begin{tabular}{|b{0.05\textwidth}|l|c|}
\hline
\centering BIC & Metabolic pathway & no. a. acids\\
\hline
\hline
\centering 34.0 & alanine, aspartate and glutamate metabolism& 6\\
\centering 33.3 & valine, leucine and isoleucine degradation& 2\\
\centering 31.3 & phenylalanine, tyrosine and tryptophan biosynthesis& 3\\
\centering 29.8 & sulfur relay system& 2\\
\centering 29.1 & glycine, serine and threonine metabolism& 7\\
\centering 27.3 & arginine and proline metabolism & 7\\
\hline
\end{tabular}
\label{tab:BIC}
\end{table*}

\section{Optimal synergy in the second order model} \label{sec:OSmodels}
\begin{figure*}[p]
\centering
\includegraphics[width=0.9\textwidth]{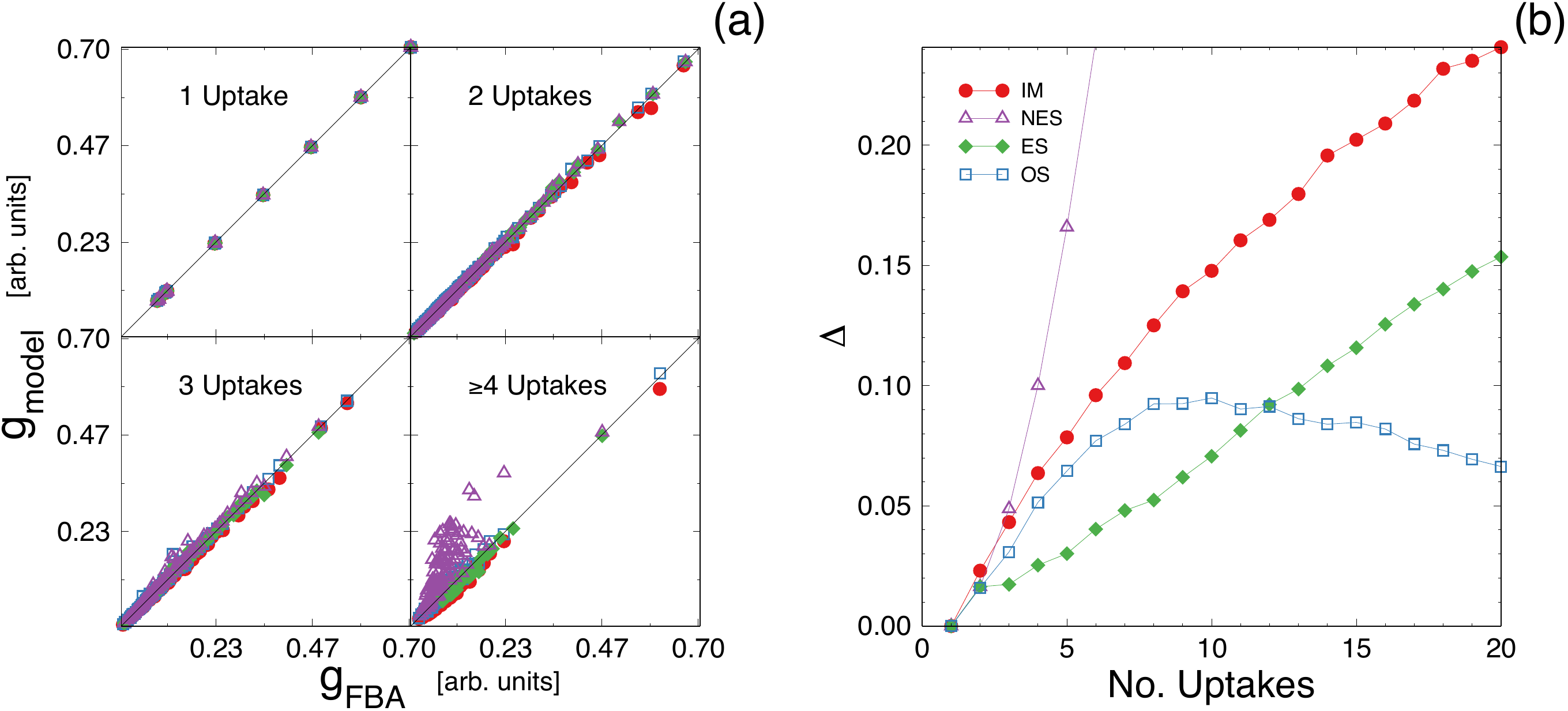}
\caption{Second order model predictions. {\bf (a)} Prediction of model bacterial growth against FBA results, for four models (see text): IM, NES, ES, \OptimizedSyn. The idealized metabolism (IM, red circles)  captures reasonably well FBA growth predictions. Including maximal synergy for all the nutrient pairs with a naive equitative synergy theory (NES, purple up triangles) largely overestimates the FBA growth. Considering a uniform uptake for all nutrient pairs with the equitative theory (ES, green diamonds) improves the IM results. When the number of uptakes is $\gg1$, all these models produce worse results than the Optimized Synergy model (\OptimizedSyn, blue squares). {\bf (b)} The relative error $\Delta$ of the different models as a function of the FBA growth $g_{\rm FBA}$. The baseline is the first order IM theory (red circles), with a relative error that increases roughly linearly with the number of uptakes. The NES model (purple up triangles) is clearly unrealistic, with a relative error that increases very fast. The ES model (green diamonds), conversely, improves the IM results, although its $\Delta$ still increases with the number of uptakes. The \OptimizedSyn~model error (blue squares) remains very low and depends very weakly on the number of uptakes, suggesting optimal allocation of synergies is a robust explanation for maximal growth.}
\label{fig:secOrds}
\end{figure*}

As shown in Sec. \ref{sec:model}, the IM model systematically underpredicts growth rates in presence of multiple nutrients. As a result we have to include a synergy term in our model. We do so by introducing the $\beta'$ functions. However, we find that an equal contribution of all synergisitc terms overpredicts the growth rate in complex media (see fig \ref{fig:secOrds}). This is because resources are limited and not all nutrient pairs can develop such maximal synergy. We therefore call this a {\em naive equitative synergy} (NES) model, that assuming maximal synergy among all nutrients describes an unrealistic scenario.

In order to limit the overall synergy, we tested the {\em equitative synergy} (ES) theory, where resources are equally distributed across the nutrient pairs. We created complex growth media as explained in Sec. \ref{sec:rmedia}, with each medium $\kappa$ consisting of $E_\kappa$ nutrients and thus $P_\kappa = E_\kappa(E_\kappa-1)/2$ possible pairs. We then assumed that, for each nutrient $i$, the uptake $\phi_i^\kappa$ was equally invested in the $E_\kappa-1$ synergies such nutrient can develop. Therefore, we computed the ES model growth on medium $\kappa$ by correcting the IM theory with the $\beta'$ contributions \eqref{eq:tanhbeta} as:
\begin{equation}
g_{\rm ES}^\kappa = g_{IM}^\kappa + \frac{1}{E_\kappa-1} \sum_{i<j} \phi_i^\kappa C_ib_{\sigma_i\sigma_j} \tanh \frac{b_{\sigma_j\sigma_i} C_j \phi_j^\kappa}{b_{\sigma_i\sigma_j} C_i \phi_i^\kappa}.
\end{equation}
Here $g_{IM}^\kappa$ is the IM theory growth, \eqref{eq:f_1_order}, $\sigma_i$ is the class of nutrient $i$, while the sum runs on the $P_\kappa$ possible nutrient pairs. Hence, with factor $1/(E_\kappa-1)$, we equally spread $\phi_i^\kappa$ across the $E_\kappa-1$ synergies.

The resulting model shows an improvement respect to the IM theory, although the gain decreases when the number of uptakes grows.

The decrease in accuracy for increasing $E$ of both the NES and the ES model suggests that the uptake of resources is distributed in some optimal way. Since in the FBA approach metabolism is aimed at growth optimization, we hypothesized that uptakes are organized in such way to maximize the nutrient synergistic contributions to growth. Specifically, such optimality must be reached by considering that nutrient uptakes that are invested to attain a certain synergy may not contribute to another synergy. In Fig. \ref{fig:all_betas}, one clearly realizes how this can be taken into account. Indeed, the $\beta'(C_{n_1}\phi_{n_1}/(C_{n_2}\phi_{n_2}))$ functions shown in Fig. \ref{fig:all_betas} typically have a growing regime followed by a plateau. The appearance of the plateau means that the synergy is not affected by a variation of the uptake of nutrient $n_1$, {\em i.e.} nutrient $n_1$ is {\em in excess} with respect to nutrient $n_2$. Conversely, in the growing region, the situation is reverted and nutrient $n_2$ is in excess. The point $T(n_1,n_2)=b_{21}/b_{12}$ marks the transition from one regime to the other. Thus, if $C_{n_1}\phi_{n_1}<C_{n_2}\phi_{n_2}b_{21}/b_{12}$, nutrient $n_2$ is in excess: in such case, $n_1$ has been completely invested and it cannot be used in other synergies, while $n_2$ can only contribute further with an effective flux $C_{n_2}\phi_{n_2}' = C_{n_2}\phi_{n_2} - C_{n_1}\phi_{n_1}T(n_1,n_2)$ \footnote{$C_{n_1}\phi_{n_1}' = C_{n_1}\phi_{n_1} - C_{n_2}\phi_{n_2}/T(n_1,n_2)$, $\phi_{n_2}=0$ respectively if nutrient $n_1$ is in excess}, that is, with the surplus of its uptake.

We hence devised the following method to achieve optimality in the case of limited resources on complex growth media:
\begin{enumerate}
\item For each pair of nutrients $i$, $j$ and corresponding uptake fluxes $\phi_i$, $\phi_j$ compute the second order correction $\Delta g_{ij}$ to the IM growth:
\begin{equation}
\Delta g_{ij} = C_j \phi_j b_{\sigma_j \sigma_i} \tanh \frac{b_{\sigma_i \sigma_j} C_i \phi_i}{b_{\sigma_j \sigma_i} C_j \phi_j},
\end{equation}
where $\sigma_i$ and $C_i$ are the class and the carbon content of nutrient $i$, respectively. 
\item Rank all $\Delta g_{ij}$ from largest to smallest. The first in such rank will be the best contribution to accomplish optimal growth.
\item \label{step:restart}Add to the IM growth prediction the first correction in the rank.
\item Reduce fluxes $\phi_i$ and $\phi_j$, so to take into account that some uptake of nutrients $i$ and $j$ has been invested into their synergy:
\begin{enumerate}
\item For the nutrient in excess, say $j$, set $\phi_j \to \phi_j - C_i/C_j\phi_ib_{\sigma_j \sigma_i}/b_{\sigma_i \sigma_j}$.
\item Set $\phi_i \to 0$, as uptake of $i$ has all been used to develop synergy $\Delta g_{ij}$.
\end{enumerate}
\item Remove from the rank all synergies involving nutrient $i$, as its effective uptake is now zero.
\item Re-compute the synergies $\{\Delta g_{kj}\}$ with the new uptake flux $\phi_j$.
\item {\em Optimal synergy} (OS) model: go to step \ref{step:restart}.
\end{enumerate}
The process is iterated until no uptake flux can be diminished further.

The above strategy to pinpoint optimal allocation of resources is really effective. The \OptimizedSyn~ model gives very accuarate results even for a large number of uptakes and we thus opted for it.

\begin{figure*}[h]
\includegraphics[width=0.9\textwidth]{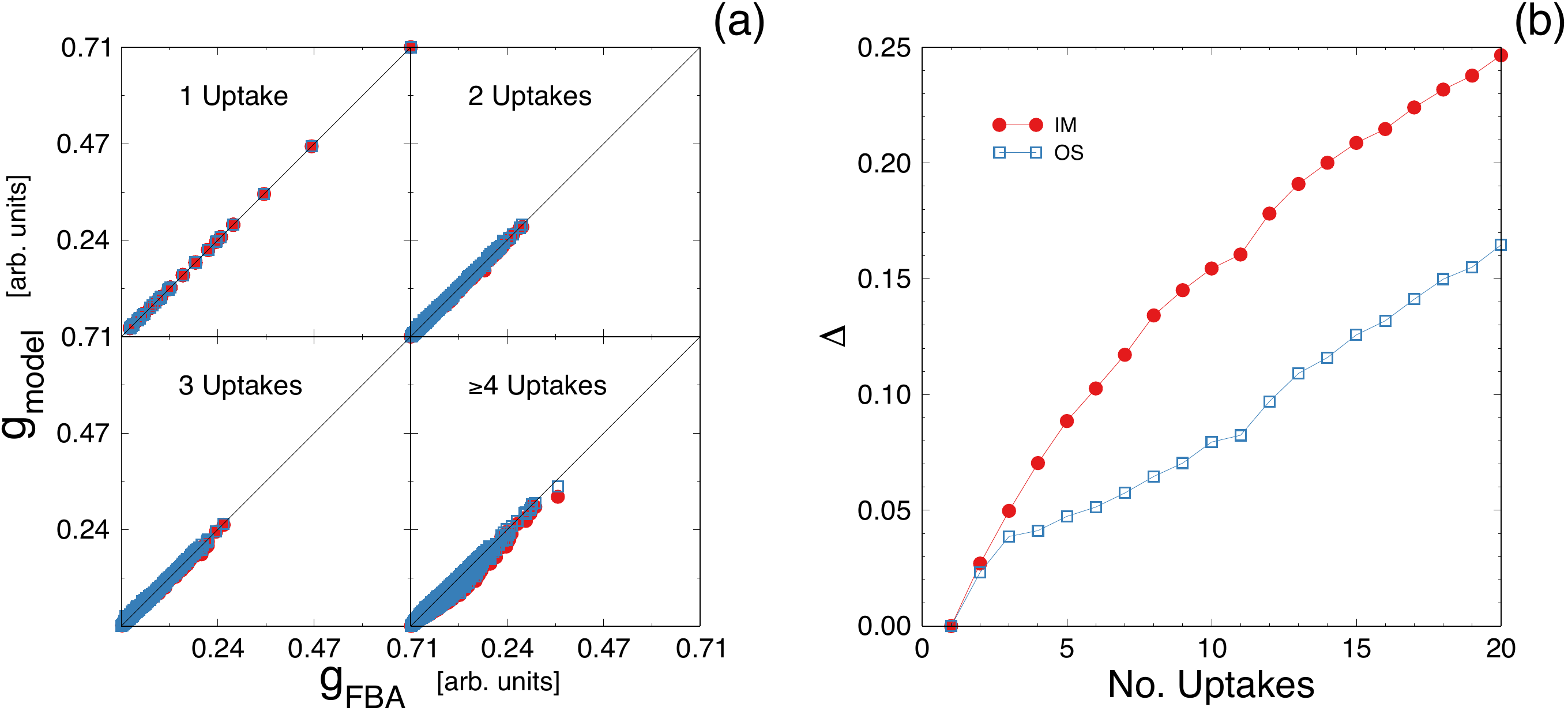}
\caption{%
{\bf (a)}  Predictions of the \OptimizedSyn~ model (blue open squares) {\em vs} the IM model (red filled circles), for complex media that may not include sugars. To better capture non--sugar synergies we allow here 2 different slopes to the $\beta$ functions %
{\bf (b)}  The relative error $\Delta$ of the \OptimizedSyn~ model (blue empty squares) and the IM model (red filled circles). Also when sugar are not always uptaken the \OptimizedSyn~model has a consistently smaller relative error than the IM model.
}
\label{fig:not_always_sugar}
\end{figure*}

Note that the results presented are derived assuming that a sugar is always present in the medium. One can generalize and also work with sugar-free complex growth media. Because $\beta'(x)$ functions for (fatty acid, base), (fatty acid, amino acid), and (amino acid, amino acid) interactions are not perfectly captured by \eqref{eq:tanhbeta} when $x\simeq 1$, this scenario is better captured allowing for two different slopes of the beta functions: results for the OS model are slightly less accurate than in presence of sugars, but still far better than the IM, as shown in Fig. \ref{fig:not_always_sugar}.
\section{Comparison with the experiments} \label{sec:ExpComps}
Beg {\em et al.}  \cite{beg07} published a few years ago a study that proves to be an excellent means to contrast our model against experimental results. In their work, the authors measured at high frequency the growth rate of a batch culture of {\em E. coli} and the corresponding variation of nutrient concentration in the medium, simultaneously. Additionally, they included in their paper measurements of the culture optical density and other quantities of interest. All the relevant measurements for our analysis are reported in Ref. \cite{beg07} Fig. 2, panels a and b: in the following, we explain how to integrate such data in our approach.

The first step to make the results of Beg {\em et al.} useful in our framework is to calculate, for each nutrient $i$, the uptakes $\phi_i$ given the time evolution of nutrient concentration $c_i(t)$ reported in Fig. 2b of Ref. \cite{beg07}. For each nutrient $i$, the uptake $\phi_i$ is related to the time derivative of the nutrient concentration $\dot{c}_i$ as:
\begin{equation}\label{eq:uptks_from_dCdt}
\phi_i(t) = V_W \frac{1}{D(t)m_i}\dot{c}_i(t),
\end{equation}
where $m_i$ is the molar mass of nutrient $i$, $D(t)$ the microbial dried mass at time $t$ and $V_W$ is the working volume, which is provided by the authors in the supporting material of Ref. \cite{beg07} (note indeed that concentration are provided per unit volume in \cite{beg07}). This relation properly yields uptakes in mmol gDW$^{-1}$ h$^{-1}$, the units commonly applied in metabolic reconstructions and that we use in our model.

From \eqref{eq:uptks_from_dCdt}, we see that, to compute $\phi_i(t)$, first the derivatives $\dot{c}_i$ must be evaluated from the provided curves $c_i(t)$, for each nutrient $i$. This is straightforward and can be accomplished with, {\em e.g.}, centered differences. For each value $\dot{c}_i(t)$ we also compute the error $\sigma_{\dot{c_i}}(t)$ evaluating the maximum and minimum slopes compatible with the given error bars of $c_i(t)$, also reported in Fig. 2b of Ref. \cite{beg07}.

The second quantity to evaluate in order to calculate the uptakes is the dried weight $D(t)$. We assume it to be proportional to the optical density $O(t)$, which is given in Fig. 2a of Ref. \cite{beg07}. Knowing the initial optical density $O(0)$ and dried weight $D(0)$ (which is specified to be $6.75\times 10^{-3}$ g), we are hence able to compute the whole $D(t)$ curve, with its own error $\sigma_D(t)$ (evaluated from the known error on the optical density).

\begin{figure*}[h]
\includegraphics[width=0.9\textwidth]{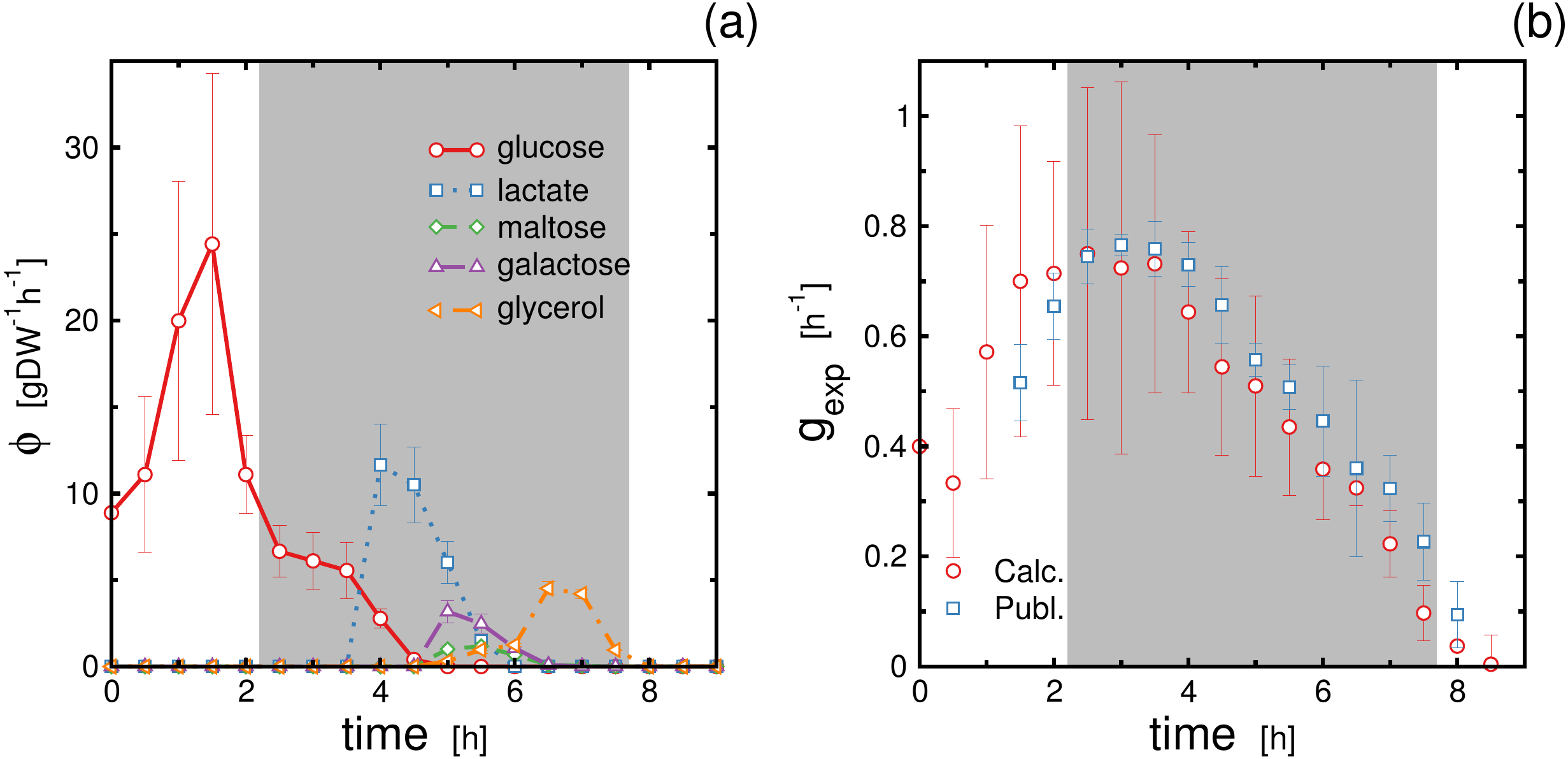}
\caption{%
{\bf (a)} The experimental uptakes $\boldsymbol{\phi}$ computed via \eqref{eq:uptks_from_dCdt}, for the five nutrients considered in Fig. 2b of Beg {\em et al.} \cite{beg07}. Glucose is almost totally consumed first, the rest of nutrients is consumed for $t> 3.5$ h. Note that the dried weight, which normalizes the plotted values, steadily grows in time. The grey shaded area is the purely exponential growth time window ($t g_{\rm expt}(t)\gtrsim 1$), where we pick the points plotted in Fig. \ref{fig:ExpComp}. %
{\bf (b)}  Comparison of the growth rate $g_{\rm expt}(t)$ calculated via \eqref{eq:Exgrate} (Calc., red circles) and the values directly published in Fig. 2a of Ref. \cite{beg07} (Publ., blue squares). The two quantities are fully consistent, all points but one being within one standard error. We use the values corresponding to the red circles to validate our model in Fig. \ref{fig:ExpComp}, as they are also related to the dried weight employed to compute the nutrient uptakes. The shaded area once again denotes the pure exponential growth region.
}
\label{fig:expData}
\end{figure*}

After the above step, we are able to compute the uptakes $\phi_i(t)$ and their associated errors $\sigma_{\phi_i}(t)$ (propagating $\sigma_{\dot{c_i}}(t)$ and $\sigma_D(t)$), for each nutrient $i$ and time $t$. Note that we do not allow negative uptakes (corresponding to nutrient release, really) and we discard noisy fluctuations of $\dot{c}_i(t)$ allowing for unexpected multiple nutrient uptakes at $t\leq 3.5$h. Consequently, $\phi_i(t)=0$ with zero uncertainty for all nutrients except glucose when $t\leq3.5$h. The resulting uptakes are plotted in Fig. \ref{fig:expData}a.

Knowing all uptakes for each time $t$, we finally compute the growth $g_{\rm OS}(t)$ predicted by the OS model by using \eqref{eq:secondorder}. We also derive an associated error $\sigma_{g_{\rm OS}}(t)$ by evaluating the growth rates yielded by the minimum $\boldsymbol{\phi}_{\rm min}(t) = \{\phi_i(t)-\sigma_{\phi_i}(t); i \in \text{nutrients}\}$ and maximum $\boldsymbol{\phi}_{\rm max}(t) = \{\phi_i(t)+\sigma_{\phi_i}(t); i \in \text{nutrients}\}$ possible uptake vectors, respectively. Therefore, in turn, $\sigma_{g_{\rm OS}}(t) = g_{\rm OS}\bigl(\boldsymbol{\phi}_{\rm max}(t)\bigr)-g_{\rm OS}\bigl(\boldsymbol{\phi}_{\rm min}(t)\bigr)$.

Albeit the experimental growth rate is partially provided in Fig. 2a of Ref. \cite{beg07}, we opt to calculate the experimental growth rate $g_{\rm expt}(t)$ resulting from our estimate of the experimental dried weight curve $D(t)$. The rationale is to have a $g_{\rm expt}(t)$ 
consistent with the $D(t)$ values used to compute the uptakes. Note indeed that in Fig. 2a of Ref. \cite{beg07} the entire time series of the experimental growth rate is not available ({\em i.e.} time window $t=0$ to $t=1.5$h is missing), so we cannot proceed the other way around and estimate $D(t)$ integrating back the growth rate. Hence, we evaluate $g_{\rm expt}(t)$ from the differential equation:
\begin{equation}\label{eq:Exgrate}
g_{\rm expt}(t) = \frac{\dot{D}(t)}{D(t)},
\end{equation}
that fixes the evolution of the dried weight in exponential growth condition. Again we estimate $\dot{D}(t)$ from $D(t)$ with centered differences and its error $\sigma_{\dot{D}}(t)$ analogously to what done for $\sigma_{\dot{c}}(t)$. Finally, we compute the error $\sigma_{g_{\rm expt}}(t)$ for $g_{\rm expt}(t)$ by propagating $\sigma_{\dot{D}}(t)$ and $\sigma_{D}(t)$. The growth rates $g_{\rm expt}(t)$ we find are entirely consistent with the ones originally published in Fig. 2a of Ref. \cite{beg07}, as shown in Fig. \ref{fig:expData}b. However, as said, such $g_{\rm expt}(t)$ values are more coherent with the dried weight we used in \eqref{eq:uptks_from_dCdt} to compute the uptakes, so these are the ones we plot in Fig. \ref{fig:ExpComp}.

\begin{figure*}
\centering
\includegraphics[width=0.6\textwidth]{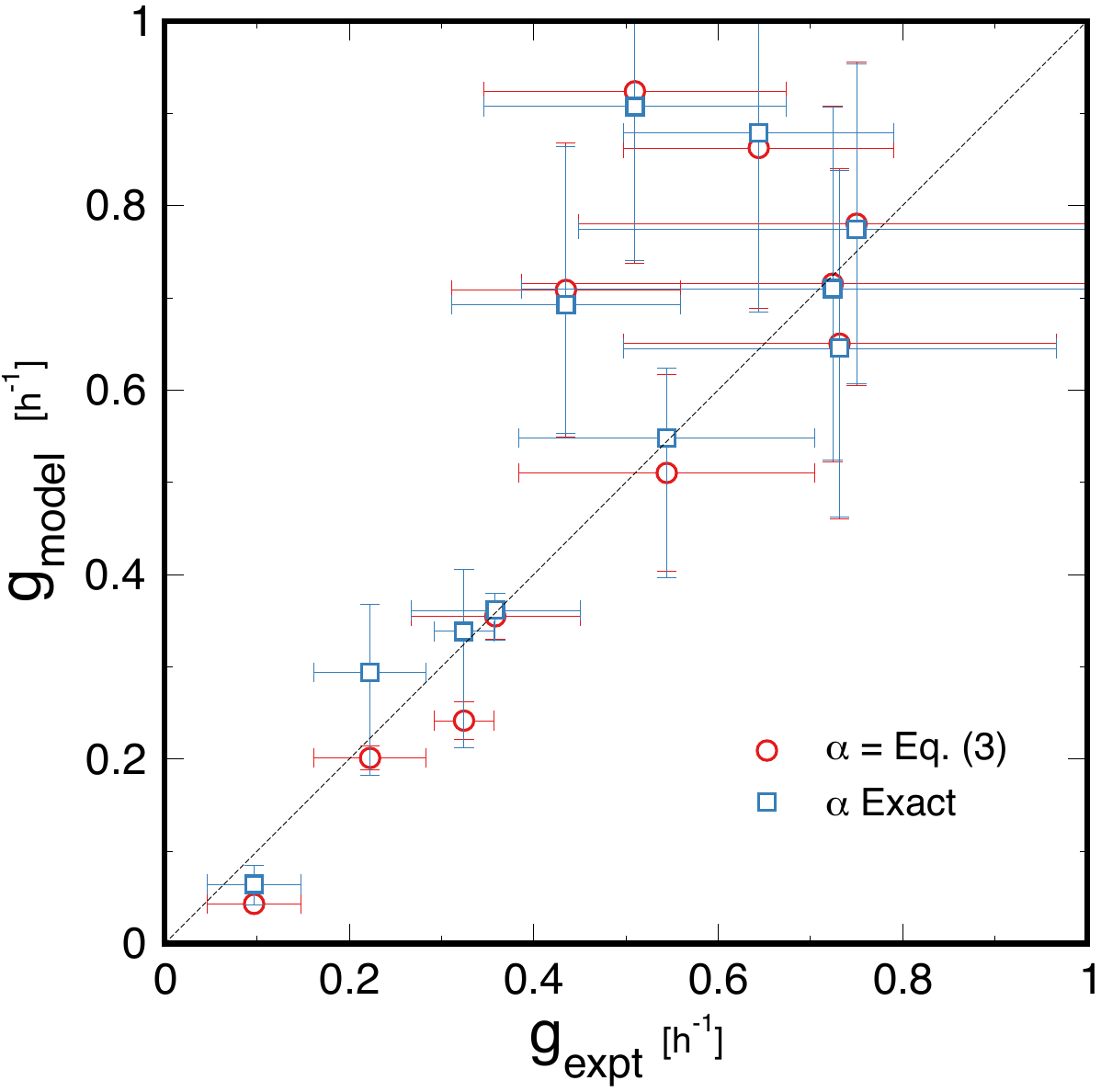}
\caption{Model prediction of experimental growthg rates. We compare here the accuracy of model \eqref{eq:secondorder} at predicting experimental bacterial growth rates when using \eqref{eq:alpha_carbons} to estimate the $\hat{\alpha}$ parameters (red circles) and by using the exact values of $\hat{\alpha}$ (blue squares), which are evaluated by estimating the nutrients yield. \eqref{eq:alpha_carbons} performs fairly well, its predictions being only slightly worse than the ones obtained with the exact $\hat{\alpha}$s. This is remarkable, as it implies that, when dealing with physiological values, one can accurately predict growth rates by only knowing the slope $a_c$ of each nutrient class and the carbon content of each nutrient, respectively, rather than the exact yield.}
\label{fig:exp_predictions}
\end{figure*}

Having computed $g_{\rm OS}(t)$ and $g_{\rm expt}(t)$, we finally compare them in Fig. \ref{fig:ExpComp}, finding an excellent agreement. To obtain these accurate results, we use \eqref{eq:alpha_carbons} to estimate the value oif each $\hat{\alpha}$. In Fig. \ref{fig:exp_predictions} we show how results change when using the exact $\hat{\alpha}$ values instead: the predictions are only slightly better. This finding is remarkable, because to use \eqref{eq:alpha_carbons} we only need to use the slopes $a_c$ (Fig. \ref{fig:alphas}b) and the carbon content of each nutrient, rather than the actual yield. The $a_c$ values hold for all nutrients in a given class, while the carbon content of nutrients is generally known, so that \eqref{eq:alpha_carbons} can be readily applied to diverse situations without having to reevaluate single nutrient contributions to growth.

Note that in these two validations against experimental results we only focus on the truly exponential growth phase, {\em i.e.} where $tg_{\rm expt}(t) \gtrsim 1$, which is the shaded region in Fig. \ref{fig:expData}.

A final remark on the fact that the experimental growth medium contains lactate and glycerol, which do not belong to nutrient classes we discuss presently. Again, one can proceed as we outline in Secs. \ref{sec:model} and \ref{subsec:2nd_order} to evaluate parameters $a$ and $b$ for the classes corresponding to these nutrients. For organic acids, the class lactate belongs to, we find $a_{\rm org\_ac}=1.5\times10^{-2}$, while $b$ parameters for all cross interactions are reported in Table \ref{tab:orgAcbeta}. For glycerol, we opt instead to use the same $a$ and $b$ parameters we derived for fatty acids, which do yield accurate results already.

\begin{table}
\caption{The OS model $b$ parameters for synergies with organic acids (${\rm org\_ac}$ in the Table). Interactions with amino acids again allows for two different plateau values of the $\beta'$ function. Nutrients are always sorted for increasing carbons: organic acids intra--class interaction does not consider pair permutations and yields thus two different $b$ values:
 $b_{\rm org\_ac~other}$ corresponds to growth on a medium where large carbon content organic acids are in excess, while $b_{{\rm other~org\_ac}}$ captures the opposite situation.}
\begin{tabular}{|l|c|c|c|}
\hline\hline
other & $b_{\rm org\_ac~other}$ &  \multicolumn{2}{c|}{$b_{{\rm other~org\_ac}}$}\\
\hline\hline
Sugars &$3.0 \times 10^{-3}$ &  \multicolumn{2}{c|}{$1.7 \times 10^{-3}$}\\\hline
Fatty acids & $3.4\times 10^{-3}$ &  \multicolumn{2}{c|}{$1.1\times 10^{-2}$}\\\hline
Organic acids & $2.7\times 10^{-3}$ &   \multicolumn{2}{c|}{$4.0\times 10^{-3}$}\\\hline
Bases & $2.4\times 10^{-3}$ &  \multicolumn{2}{c|}{$2.4\times 10^{-2}$}\\\hline
Amino acids &$3.0\times 10^{-3}$ & $3.0\times 10^{-3}$ &$1.8\times 10^{-2}$\\
\hline
\end{tabular}

\label{tab:orgAcbeta}
\end{table}

\bibliographystyle{apsrev4-1}
\bibliography{MetabolismScaling_main_lite}
\end{document}